\documentclass[letterpaper,journal]{IEEEtran}
\usepackage{amsmath,amsfonts}
\usepackage{algorithmic}
\usepackage{algorithm}
\usepackage{array}
\usepackage[caption=false,font=normalsize,labelfont=sf,textfont=sf]{subfig}
\usepackage{textcomp}
\usepackage{stfloats}
\usepackage{url}
\usepackage{verbatim}
\usepackage{graphicx}
\usepackage{soul}
\usepackage{xcolor}
\usepackage{cite}

\usepackage{color}
\usepackage{soul}

\begin{document}

\title{Fast-Convergent Meta-RL via Gradient-Clustered BS Sampling for Edge Caching}

\author{Farnaz Niknia,
Ping Wang,~\IEEEmembership{Fellow, IEEE}

\thanks{}

\thanks{Farnaz Niknia and Ping Wang are with the Lassonde School of Engineering
at York University, Toronto, ON, M3J 1P3, Canada
(email: fniknia@yorku.ca, pingw@yorku.ca)}
}

\maketitle

\begin{abstract}
Wireless edge caching networks typically consist of many independent
Base Stations (BSs), each facing its own request rate and
content popularity profile. Training a Reinforcement Learning
(RL) caching agent from scratch at every BS forces each agent
to relearn, through slow trial and error, a decision problem
that is structurally identical across the network.
Meta-reinforcement learning removes this redundancy by
learning a shared initialization that adapts to any BS in a
few local updates; however, meta-training itself becomes the
bottleneck at scale: the meta-gradient must be estimated from
a small subset of BSs at each meta-iteration, and sampling
this subset uniformly at random yields a high-variance
estimate, an issue existing meta-RL caching
frameworks leave unaddressed. This paper proposes a
meta-reinforcement learning framework for caching across
independent, non-overlapping BSs that directly targets this
bottleneck. Each BS runs a local Proximal Policy Optimization
(PPO) agent, formulated as a Semi-Markov Decision Process (SMDP) over
content popularity, size, lifetime, and importance, while a
shared meta-policy is learned via a Model-Agnostic
Meta-Learning (MAML)-style loop. To scale meta-training and accelerate convergence, we
introduce gradient-based clustering, which groups BSs by local
gradient similarity and draws from every cluster, in
proportion to its size, at each meta-iteration. We prove, via
an Analysis of Variance (ANOVA)-style decomposition of gradient variance, that this
strategy yields a strictly lower-variance meta-gradient estimator than uniform random sampling under BS heterogeneity.

\end{abstract}


\section{Introduction}
\label{Introduction}

\IEEEPARstart{T}{he} continued expansion of mobile applications and
connected services has placed mounting strain on backhaul
links and core networks, as billions of devices generate an
ever-growing volume of content requests that must ultimately
be served from centralized data centers. Video streaming
platforms, cloud-based applications, and the proliferation of
Internet of Things (IoT) devices have driven this surge in
demand, motivating network operators to push content storage
closer to end users through edge caching. By storing
frequently requested content at BSs near the
network edge, operators can substantially reduce backhaul
load, shorten response times, and improve the overall quality
of experience delivered to users.

A practical edge caching deployment, however, rarely consists
of a single cache node. Operators typically deploy many BSs
across a coverage region, each independently serving its own
local population of users with its own request rate and
content popularity profile~\cite{somesula2024cooperative,
feng2025federated, qian2025federated}. A BS situated in a
dense urban core, for instance, may experience high request
volumes concentrated around a narrow set of trending content,
while a BS in a lightly populated suburban area may see
sparser, more evenly distributed requests. Training a
RL caching agent independently from
scratch at every one of these BSs is wasteful: although the
traffic statistics differ from one BS to another, the
underlying caching decision problem, namely which files to
retain given memory constraints, content attributes, and
request patterns, is structurally identical across all BSs.
This redundancy suggests that knowledge learned at one BS
should, in principle, be transferable to accelerate learning
at another, yet most prior multi-node caching frameworks
either assume tight coordination between BSs through explicit
cooperation or federated aggregation~\cite{somesula2024cooperative,
feng2025federated}, or treat each BS as fully isolated with no
mechanism for knowledge transfer at all \cite{zhou2023edge, zhong2020deep}.

Meta-reinforcement learning offers a principled mathematical
framework that explicitly optimizes for fast adaptation
to new BSs after minimal local updates, rather than simply
transferring knowledge as a fixed initialization. The key
distinction is fundamental: transfer learning treats the
pre-trained policy as a one-time starting point; in contrast,
meta-learning solves an outer-loop optimization problem that
seeks an initialization $\theta$ such that just a few inner-loop
gradient steps on a new BS's local data yield a
high-performing policy for that BS. This is formalized as
minimizing the post-adaptation loss averaged over all training
BSs, which directly incentivizes rapid
convergence at new BSs rather than converging quickly to a
single fixed policy. Moreover, meta-learning provides formal
convergence guarantees under the Model-Agnostic Meta-Learning (MAML) framework \cite{finn2017model}. In practical
terms, this means a meta-trained initialization can adapt to
a new BS's traffic conditions, , such as sudden shifts in content popularity or changes in request rates, with only a small number of local updates while maintaining or
improving performance. A transfer-learned policy, by contrast,
offers no such guarantee and often requires retraining or
careful hyperparameter tuning to avoid negative transfer when
the new environment deviates significantly from that observed during training. We exploit this observation through
meta-reinforcement learning, specifically following the
MAML framework~\cite{finn2017model},
which enables fast adaptation to new and unseen BS
environments while maintaining theoretical convergence
guarantees.

Realizing this benefit at scale, however, introduces a design
challenge that has received little attention in the existing
literature. Computing the meta-gradient that drives the
outer-loop update of the shared policy requires evaluating
local adaptation steps across BSs, and as the number of BSs in
the network grows, evaluating every BS at every meta-iteration
becomes computationally prohibitive. The standard remedy is to
estimate the meta-gradient from a randomly sampled subset of
BSs at each iteration. This is where uniform random sampling
becomes problematic: when BSs differ substantially in their
traffic characteristics, a randomly drawn subset may by chance
overrepresent BSs from one traffic regime while leaving others
entirely unsampled, producing a meta-gradient estimate that is
unrepresentative of the network as a whole and that varies
considerably from one meta-iteration to the next. This
high-variance estimation directly slows the convergence of the
shared policy, yet, most existing work on meta-RL for edge caching relies on uniform random sampling or assumes networks small enough that all BSs can participate in every meta-iteration, leaving the important question of how to effectively select a subset of BSs at each meta-iteration largely unexplored.

To address this limitation, we introduce a meta-reinforcement
learning framework for caching across a network of
independent, non-overlapping BSs, in which each BS executes a
local Proximal Policy Optimization (PPO) ~\cite{schulman2017proximal}
agent that accounts for content popularity, request arrival rate together with key
file attributes, including file size, lifetime, and importance, and a
shared meta-policy is learned across BSs using a
MAML-style outer
loop~\cite{finn2017model}. To make this meta-learning process
scale efficiently to large BS populations and accelerate convergence, we propose a
gradient-based clustering mechanism that periodically clusters
BSs according to the similarity of their most recently
computed local gradients and selects BSs from each cluster proportionally to the cluster size when constructing the BS batch at each meta-iteration. We
further provide a formal convergence analysis, grounded in an
Analysis of Variance (ANOVA)-style decomposition of gradient variance into
within-cluster and between-cluster components, proving that
this gradient-clustered sampling strategy yields a
strictly lower-variance meta-gradient estimator than uniform
random sampling whenever BSs are heterogeneous in their
traffic characteristics, and that this variance reduction
translates directly into a faster expected decrease of the
meta-loss over training.

The
contributions of this paper are summarized as follows.

\begin{itemize}

  \item{We propose a meta-reinforcement learning framework,
  built on a MAML-style inner-outer loop structure, that
  learns a shared policy initialization across all BSs in the
  network, enabling any individual BS to adapt to its local
  traffic conditions, or to a shift in those conditions, using
  only a small number of local gradient updates rather than
  training a caching policy from scratch.}

  \item{We introduce a gradient-based clustering mechanism for
  selecting the BS batch used to estimate the meta-gradient
  at each meta-iteration, in which BSs are periodically clustered
  by the similarity of their local gradients and BSs
  are sampled from each cluster proportional to its size, ensuring that every traffic
  regime present in the network is represented in every
  meta-iteration.}

  \item{We prove, via an ANOVA-style decomposition of gradient
  variance into within-cluster and between-cluster components,
  that gradient-clustered BS sampling yields a 
  lower-variance meta-gradient estimator than uniform random
  sampling whenever BSs are traffic-heterogeneous. Alao, the known convergence guarantee for
  MAML~\cite{fallah2020convergence} still holds under our
  sampling scheme, with the total gradient variance replaced
  by its smaller within-cluster part.}

  \item{{We evaluate the proposed approach through extensive simulations on a heterogeneous network of independent BSs, and show that it (i) empirically confirms the predicted variance decomposition and reduction, (ii) converges  faster and more stably than uniform random sampling during meta-training, and (iii) matches the strongest baselines in adapted caching performance while avoiding the negative transfer that degrades transfer learning, thereby validating our theoretical results.}}

\end{itemize}

The remainder of this paper is as follows. Section ~\ref{Related Work} reviews related work and section~\ref{sec:sysmodel}
introduces our multi-BS system model, while
Section~\ref{sec:formulation} formalizes the per-BS caching
problem. Section~\ref{sec:algorithm} presents our
proposed meta-RL caching algorithm with gradient-clustered
BS sampling, together with the accompanying convergence
analysis. Section~\ref{results} describes our experimental
setup and results, and Section~\ref{Conclusion} concludes the
paper. For reference, Table~\ref{table:notation} provides a
comprehensive summary of all notations used throughout the
paper.

\renewcommand{\arraystretch}{1.4}
\begin{table}[htbp]
  \centering
  \caption{Table of notations}
  \label{table:notation}
  \begin{tabular}{lp{0.7\linewidth}}
  \hline
    \textbf{Symbol} & \textbf{Description} \\
    \hline
    
    $\lambda_b$ & Poisson request arrival rate at BS $b$ \\
    
    $\mathcal{F}$ & Content catalog,
    $\mathcal{F} = \{f_1, f_2, \ldots, f_F\}$ \\
    
    $\mathbf{d}(t)$ & Popularity vector at time $t$ \\
    
    $l_f$ & Lifetime (validity duration) of content type $f$ \\
    
    $z_f$ & Size (in storage units) of content type $f$ \\
    
    $i_f$ & Importance (priority weight) of content type $f$ \\
    
    $h^f(t)$ & Freshness of content type $f$ at time $t$, $h^f(t) \in [0,1]$ \\
    
    $y^f(t)$ & Utility of content type $f$ at time $t$ \\

    $M$ & Cache capacity at each BS (in storage units) \\
    
    $a_b(t)$ & Cache action at BS $b$: $a_b(t) \in \{0,1\}$ \\
    
    $\tau_b$ &inter-interval time between requests at BS $b$ \\
    
    $r_b(t)$ & Instantaneous reward collected by BS $b$ at decision epoch $t$ \\
    
    $\gamma$ & Discount factor \\

    $\pi_b$ & Caching policy at BS $b$ \\
    
    $\pi_b^*$ & Optimal stationary policy at BS $b$ \\
    
    $\theta$ & Shared meta-policy parameters \\
    
    $\theta_b$ & BS-specific policy network parameters \\
    
    $\theta_b'$ & Adapted policy parameters after inner-loop update at BS $b$ \\
    
    $\alpha$ & Inner-loop (BS-specific) learning rate \\
    
    $\eta$ & Outer-loop (meta) learning rate \\
    
    $g_b(\theta)$ & Local gradient at BS $b$ \\
    
    $\hat{G}(\theta)$ & Meta-gradient estimator (sampled over BS batch $S$) \\
    
    $G^*(\theta)$ & True population meta-gradient (over all BSs $\mathcal{B}$) \\
    
    $S$ & BS batch (subset of BSs sampled for meta-iteration) \\
    
    $K$ & Number of clusters \\
    
    $m$ & Sampling budget (BS evaluations per meta-iteration), $m \ll N$ \\
    
    $m_k$ & Number of BSs drawn from cluster $C_k$: $m_k = m\,n_k/N$ \\
    
    $N$ & Total number of BSs, $N = |\mathcal{B}|$ \\
    
    $C_1, \ldots, C_K$ & Clusters of BSs based on gradient similarity \\
    
    $n_k$ & Size of cluster $C_k$ \\
    
    $\mu_k$ & Mean gradient within cluster $C_k$ \\
    
    $\sigma_k^2$ & Within-cluster gradient variance for cluster $C_k$ \\

    $\sigma^2$ & Total gradient variance\\
    
    $\Delta_c$ & Re-clustering interval \\
   
    \hline
  \end{tabular}
\end{table}

\section{Related Work}
\label{Related Work}

Research on edge caching with reinforcement learning can be
organized along three threads relevant to this work:
single-node Deep Reinforcement Learning (DRL)-based caching, federated
caching across base stations, and meta-learning approaches. Next, we review each in turn.

\subsection{Single-Node Edge Caching}

DRL has been widely adopted for caching at a
single edge node, where an agent observes local request dynamics and learns
a policy that maximizes cache efficiency without prior knowledge of the
environment. The authors in \cite{zhu2019transient} model the caching of transient
data at an edge router as a discrete-time Markov Decision Process (MDP) and
develop an actor--critic DRL method that balances communication cost against
data freshness. Similarly, the work in \cite{wu2022transient} employs PPO to jointly improve the cache hit rate and reduce
energy consumption while accounting for the limited lifetime of IoT data.
Building on this line of research, the study in \cite{zhang2024tdmeac}
proposes TD-MEAC, a maximum-entropy actor--critic strategy for transient
data that constructs an explicit freshness model and achieves higher hit
rates and average freshness than existing DRL-based schemes. 

More recent efforts have incorporated attention and Transformer
architectures into learning-based caching at a single node. T-CacheNet
\cite{kim2024tcachenet} employs a Transformer-decoder-based DRL framework in
which all users request content from a single edge server with limited cache
capacity, using self-attention to encode request patterns. The authors in
\cite{kim2025miss} formulate cache replacement as a partially observable MDP
and propose the miss-triggered Cache Transformer (MTCT), a
Transformer-decoder Q-learning agent that encodes recent request histories
with self-attention and invokes its policy only upon cache misses. In
\cite{teng2024attention}, an attention mechanism aids a DRL agent in jointly
optimizing proactive caching and cache replacement in a mobile edge
computing network under a cloud--edge--device architecture.

\subsection{Federated Edge Caching}

A second line of research distributes the learning process across multiple
caching nodes, using federated learning to share model parameters rather
than raw request traces. The FADE framework in \cite{wang2020fade} pioneers
this direction, letting each edge node train a local DRL caching agent and
periodically upload its weights to a central server for aggregation, thereby
improving hit rate and offloading backhaul traffic without exposing user
data. To account for the differing content preferences of neighboring nodes,
the authors of \cite{wang2021attention} propose an attention-weighted
federated DRL scheme for device-to-device assisted collaborative caching, in
which attention weights determine each participant's contribution to the
aggregated model. Study in \cite{sun2023recommendation} integrates recommender
systems with edge caching in a two-tier edge--cloud network and adopts a
multi-agent actor--critic algorithm trained under federated learning,
enabling edge servers to independently learn caching strategies that
minimize long-term system cost. In a related effort, \cite{zhou2024federated}
models recommendation-enabled caching as a partially observable MDP, applies
deep deterministic policy gradient to learn the policy, and employs an Earth
mover's distance criterion for personalized model aggregation.

Subsequent work addresses the communication overhead and heterogeneity that
federated training introduces. The CE-FDRL method of \cite{zhang2024ce}
prunes and quantizes the shared DRL models and lengthens the aggregation
interval to reduce the number of transmitted parameters, and provides a
global convergence analysis for cooperative caching in fog radio access
networks. Similarly, the study in \cite{wu2024elastic} combines elastic federated
learning with multi-agent DRL so that fog access points with unequal
computational budgets can participate in cooperative caching. Closer to the
question of environment heterogeneity, the authors of \cite{li2024pfdrlca} observe that a
single aggregated model adapts poorly to edge servers with dissimilar
request distributions, and propose a personalized federated training
framework in which a Multi-Head Deep Q-Network (MH-DQN) is trained
layer-wise, so that shared layers capture global structure while
personalized layers specialize to each local environment.

Federated approaches thus establish that knowledge acquired at one caching
node can usefully inform another. However, their objective is to fit a set
of policies to the base stations observed during training: aggregation is
tied to the participating nodes, and a newly deployed base station, or an
existing one whose request statistics have drifted, must either accept a
global model that was never optimized for its own dynamics or undergo a
fresh round of federated training. Furthermore, personalization is achieved by partitioning network layers rather
than by optimizing explicitly for adaptability. By contrast, the
meta-learning formulation adopted in this paper treats each base station as
a task drawn from an underlying task distribution and optimizes the shared
initialization so that, by construction, a small number of gradient steps on
a previously unseen task recovers a near-optimal policy. Federated learning
and meta-learning are therefore complementary rather than competing: the
former addresses where computation and data reside, whereas the latter
addresses how quickly a policy can adapt to a task it has never seen.


\subsection{Meta-Reinforcement Learning for Edge Caching}

Several works apply meta-RL to caching at a \emph{single} node. The
scheme in \cite{sakr2023meta} pairs a Deep Deterministic Policy Gradient (DDPG) base learner with a MAML
meta-learner for proactive caching in a vehicular network served by one
road-side unit. Each BS corresponds to a different Zipf popularity setting, and the meta-learner is trained across these BSs so that the shared initialization can adapt quickly to a new one. Since only the popularity skewness changes from task to task, the request timing is never part of what makes two tasks different.  The MRFAC algorithm
in \cite{ye2026fast} likewise models single-node mobile edge caching as an
MDP and adds a meta-learner whose gated-recurrent-unit weights encode the
prior knowledge transferred to a new BS; tasks are again induced by shifts
in the request pattern at one node, and the adaptation performance is evaluated against baselines Least Frequently Used (LFU), First-In First-Out (FIFO), and DDPG. In both cases the model is discrete-time, so the request
arrival rate plays no part in what separates one task from another.

Three works \cite{mao2023collaborative}, \cite{alshedivat2018continuous} and \cite{wei2024cooperative} operate on multiple nodes and are therefore closer to the present setting; they differ mainly in how tasks and training BS batches are formed. In \cite{mao2023collaborative}, a task is the request pattern seen at one edge during a short time window. Following~\cite{alshedivat2018continuous}, the method pairs consecutive time windows and learns to adapt from the earlier to the later one, using both same-node and neighbor-to-local pairs. Rather than sampling a subset of BSs, it uses all available pairs; the only selection is over neighbors during adaptation, where a learned weight matrix upweights those most similar to the local node. The approach in \cite{wei2024cooperative} is a meta-RL
scheme in which an inner DDPG model lets several base stations with
overlapping coverage cooperatively decide cache replacement, while an outer
MAML-style model learns an initialization that generalizes to unseen
popularity distributions. Here, tasks differ in their content-popularity distribution, and, unlike \cite{mao2023collaborative}, each
meta-iteration randomly samples a batch of BSs and averages the resulting parameter differences to update the
meta-parameters. Training and test BSs are drawn from disjoint Zipf
skewness ranges to test generalization to unseen popularity.

\subsection{Motivation}
Two observations follow. First, across all of the aforementioned works, a ``task'' is
induced by varying the popularity skewness while the temporal statistics of
the request process are held fixed; because the underlying models are
discrete-time, the request rate and hence the interarrival distribution that
determines how long a cached file survives before it expires, cannot
distinguish one task from another. To address this limitation, we adopt the continuous-time SMDP formulation in this work, and
 base stations with identical popularity profiles but
different arrival rates are genuinely different tasks. Second, existing meta-RL
caching methods either enumerate all task pairs \cite{mao2023collaborative},
draw BSs at random \cite{wei2024cooperative}, or partition them by a
hand-chosen popularity parameter \cite{sakr2023meta,ye2026fast}. The framework developed in this paper
addresses both points: (i) tasks correspond to base stations with heterogeneous
popularity and arrival dynamics, (ii) the BSs are sampled using gradient-based clustering rather than random or hand-partitioned sampling, and
the resulting meta-policy is theoretically shown to converge faster than its counterpart under random sampling.

\section{System Model}
\label{sec:sysmodel}

\subsection{Network Architecture}

\begin{figure}[!t]
  \centering
  \includegraphics[width=\linewidth]{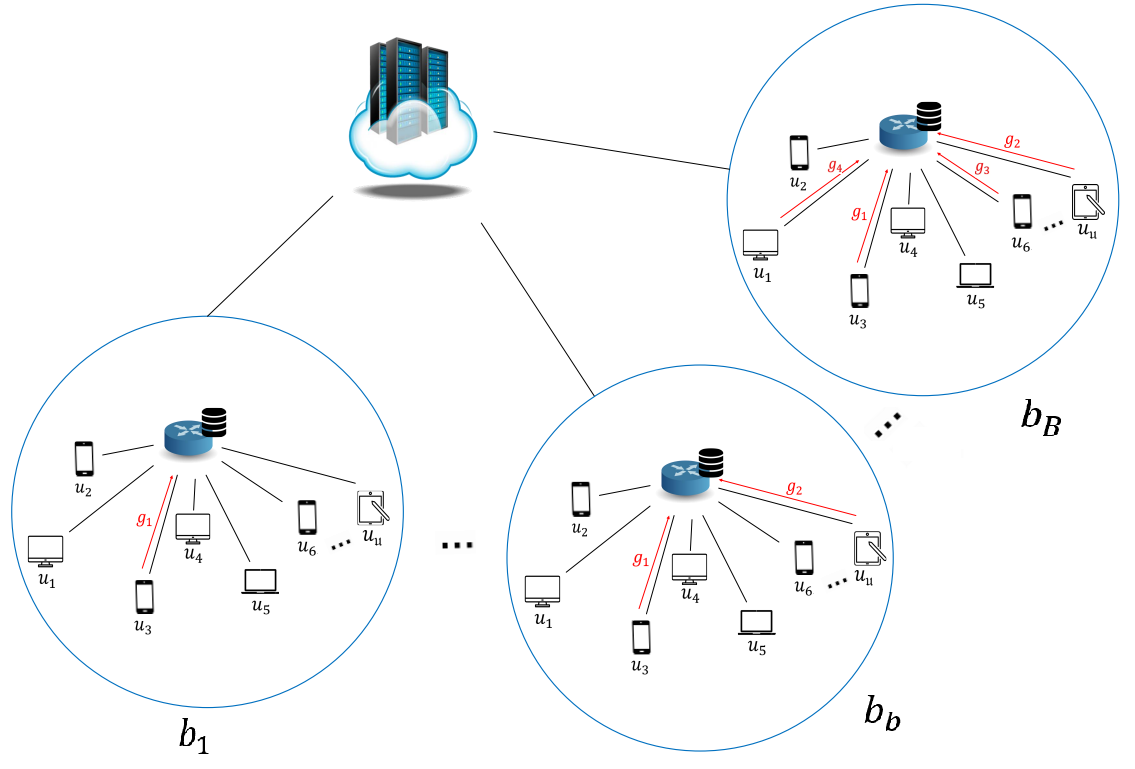}
  \caption{System model: a central server and $B$ independent, non-overlapping BSs, each serving its own users from a local cache
  and characterized by its own request rate $\lambda_b$ and
  content popularity distribution $\mathbf{d}_b(t)$. The BSs do
  not share cache contents or coordinate their caching
  decisions.}
  \label{fig:sysmodel}
\end{figure}

We consider a heterogeneous wireless network comprising
$N$ BSs, indexed by
$\mathcal{B} = \{1, 2, \ldots, N\}$, all connected to a
central content server, as illustrated in Fig. \ref{fig:sysmodel}.
Each BS $b \in \mathcal{B}$ serves a  set of
end-user devices within its own coverage area, and
neighboring BSs have non-overlapping service zones so that
every user is associated with exactly one BS.
Because the BSs operate independently of one another,
there is no inter-BS coordination or content sharing;
each BS makes its own caching decisions based solely on
the requests generated by its local users.

Each BS is equipped with an embedded cache of capacity
$M$ (in storage units).
The central server is assumed to store the complete
content catalog and has sufficient capacity to respond to
any request forwarded by the BSs~\cite{wei2021wireless}.
When a user attached to BS $b$ issues a request, the BS
first consults its local cache. A \emph{cache hit} occurs
when the requested item is found locally and is served
immediately, avoiding a round trip to the server.
A \emph{cache miss} triggers a fetch from the central
server, incurring additional transmission delay and
consuming backhaul bandwidth.
Hereafter, the term \emph{cache} refers to the storage
unit collocated with a BS.

Since the BSs are statistically independent (disjoint
user populations, no shared cache), the caching problem
at each BS is self-contained.
Accordingly, the system model and problem formulation that
follow are stated for a representative BS $b$; an
identical formulation applies to every other BS in
$\mathcal{B}$.

Let $\mathcal{U}_b = \{u_1, u_2, \ldots, u_{U_b}\}$
denote the set of users served by BS $b$.
Requests generated within the coverage area of BS $b$
arrive in sequence and are indexed by
$\mathcal{G}_b = \{g_1, g_2, \ldots, g_{G_b}\}$,
where $g_g$ identifies the $g$-th request irrespective
of the originating user.
Requests are processed in arrival order
(first-come, first-served).

The content catalog shared across all BSs is denoted by
$\mathcal{F} = \{f_1, f_2, \ldots, f_F\}$,
where $f_f$ represents the $f$-th content type.
Content items originate from diverse sources (cameras,
sensors, and edge computing nodes) and are stored at the
central server.
Each content type carries four application-specific
attributes, defined below.

\subsubsection*{Content Attributes}
\begin{enumerate}
  \item \textit{Popularity ($d_f$):}
        The number of times content type $f$ is requested
        within a predefined observation window $B$ at a
        given BS~\cite{li2016popularity}.
        Items with high request frequency (such as
        trending video clips) are considered popular and
        are stronger candidates for local caching.

  \item \textit{Lifetime ($l_f$):}
        The duration for which content type $f$ remains
        valid after leaving the server~\cite{nasehzadeh2020deep}.
        Time-critical data (e.g., real-time sensor
        readings or location updates) expires quickly,
        whereas archival content may remain relevant
        indefinitely~\cite{vural2017caching}.

  \item \textit{Size ($z_f$):}
        The amount of cache storage consumed by one
        instance of content type $f$.
        Large items (e.g., high-definition video) compete
        more aggressively for the limited cache capacity
        $M$ of a BS than compact items such as text files.

  \item \textit{Importance ($i_f$):}
        A scalar reflecting the priority users and
        operators assign to content type $f$.
        Mission-critical or safety-relevant content
        receives a high importance score, while
        low-priority entertainment content receives a
        lower score.
\end{enumerate}

A sound caching policy must account for all four
attributes jointly.
For instance, a content item that is highly popular but
large, short-lived, and unimportant may contribute less
to overall system performance than a moderately popular
item that is compact, long-lived, and important.

The popularity vector observed at BS $b$ at time $t$ is
$\mathbf{d}(t) = (d_1(t), d_2(t), \ldots, d_F(t))$,
where $d_f(t)$ records how often content type $f$ was
requested during window $B$ at that BS.
The lifetime, size, and importance vectors are written as
$\mathbf{l} = (l_1, \ldots, l_F)$,
$\mathbf{z} = (z_1, \ldots, z_F)$, and
$\mathbf{i} = (i_1, \ldots, i_F)$, respectively,
and are treated as global attributes shared across all
BSs.

\subsubsection*{Content Utility}
The benefit of retaining a content item in the cache of a
BS is jointly determined by how fresh the item is and how
important it is to the local user population.
The \emph{freshness} of content type $f$ at time $t$ is
defined as
\begin{equation}
  h^f(t) \;=\; \left(\frac{t - w^f_g}{l_f}\right),
  \qquad 0 \leq h^f(t) \leq 1,
\end{equation}
where $w^f_g$ is the time at which content type $f$ was
generated and $l_f$ is its lifetime \cite{niknia2025attention}.
The freshness value is zero when the item is generated and increases as the item ages, reaching one upon expiration.

The utility of content type $f$ cached at BS $b$
at time $t$ is expressed as
\begin{equation}
  y^f(t) \;=\; \mathcal{E}\!\left(h^ f(t),\; i_f\right),
  \label{ufunc}
\end{equation}
where $\mathcal{E}(\cdot)$ an increasing function with respect to $i_f$ and is a decreasing function with respect to $h^f(t)$ \cite{niknia2025attention}.

\subsection{Environment Uncertainties}

Each BS independently faces two key sources of
environmental uncertainty that can degrade caching
performance if left unaddressed.

\subsubsection{Content Popularity Dynamics}
The popularity distribution observed at a BS is
non-stationary.
User demand shifts in response to social trends,
time-of-day effects, and unpredictable events such as
breaking news or viral content, causing certain items to
surge in request frequency while others rapidly lose
traction.
When the popularity distribution changes, a caching
policy optimized for prior conditions may retain items
that are no longer in demand while failing to cache newly
popular ones, leading to a degraded cache-hit rate.

\subsubsection{Request-Rate Variability}
In addition to changes in which content is popular,
the \emph{rate} at which requests arrive at a BS also
fluctuates over time.
Factors such as user density variations, device
connectivity patterns, and time-of-day traffic cycles
all introduce variability in inter-request intervals.
Request arrivals at each BS are modeled as an
independent Poisson process with rate $\lambda_b$, a
standard assumption for characterizing wireless user
traffic~\cite{gomaa2013estimating}.
Since BSs serve disjoint user populations, the arrival
processes at different BSs are mutually independent.
Each BS has no prior knowledge of its own $\lambda_b$ or
of any parameters of the governing Poisson process.
A caching policy tuned to a particular arrival rate may
either over-replace cached items during periods of low
demand or fail to keep pace with high-demand bursts,
resulting in suboptimal utilization of the available
cache capacity $M$.
Consequently, monitoring and reacting to shifts in
$\lambda_b$ is as critical as adapting to changes in the
local popularity distribution.

\section{Problem Formulation}
\label{sec:formulation}

The caching problem at each BS is inherently sequential:
the BS must decide, upon every content request, whether
to store the retrieved item locally, taking into account
the current cache occupancy, the attributes of the
incoming content, and the long-term effect of each
decision on future performance.
This structure maps naturally onto a
SMDP, which
generalizes the standard MDP to settings where the time
elapsed between consecutive decision epochs is itself a
random variable~\cite{puterman2014markov}.
We adopt the SMDP framework because content requests
arrive according to a Poisson process, making transition
times non-uniform and ruling out the fixed unit-step
assumption of a discrete-time MDP.

Since all BSs are statistically independent (disjoint
coverage areas, no shared cache, no inter-BS
collaboration), the SMDP at each BS $b$ is
self-contained and structurally identical.
The formulation below therefore describes a single
representative BS $b$; the same model applies to every
other BS in $\mathcal{B}$ without modification.

The SMDP at BS $b$ is defined by the five-tuple
$(\mathcal{S}, \mathcal{A}, \mathcal{J}, \mathcal{R}, \pi)$,
where $\mathcal{S}$ is the state space,
$\mathcal{A}$ is the action space,
$\mathcal{J}$ captures state transition probability,
$\mathcal{R}$ is the reward function, and
$\pi$ denotes the caching policy.
Each component is described in detail below.

\subsubsection{State Space}

At each decision epoch $t$, the state of BS $b$ is
\begin{equation}
  s_b(t) \;=\;
  \bigl\{\,
    \mathrm{Mem}_b(t),\;
    \mathbf{b}_b(t),\;
    \mathbf{y}_b(t),\;
    \mathbf{d}_b(t),\;
    \mathbf{i}(t),\;
    \mathbf{l}(t),\;
    \mathbf{z}(t)
  \,\bigr\},
\end{equation}
where the components are defined as follows.

$\mathbf{b}_b(t) \in \{0,1\}^F$ is a binary occupancy
vector whose $f$-th entry equals $1$ if content type $f$
is currently stored in the cache of BS $b$ and $0$
otherwise.

$\mathrm{Mem}_b(t)$ is the normalized available cache
capacity at BS $b$, given by
\begin{equation}
  \mathrm{Mem}_b(t)
  \;=\;
  \frac{M \;-\; \sum_{f=1}^{F} b^f_b(t)\, z_f}{M},
\end{equation}
where $M$ is the cache capacity of each BS and $z_f$ is
the size of content type $f$.
A value of $\mathrm{Mem}_b(t) = 0$ indicates a
fully occupied cache, while $\mathrm{Mem}_b(t) = 1$
corresponds to an empty cache.

$\mathbf{y}_b(t)$ tracks the utility of each item
currently cached at BS $b$.
Specifically, $y^f_b(t) = 0$ for any content type $f$
not present in the cache, and $y^f_b(t) > 0$ otherwise,
with its value determined by the utility function in Eq.~\ref{ufunc}.

\subsubsection{Action Space}

At every decision epoch $t$, the caching agent at BS $b$
chooses one of two actions for the content item associated
with the arriving request:
\begin{equation}
  a_b(t) \;\in\; \{0,\, 1\},
\end{equation}
where $a_b(t) = 1$ instructs the BS to store the item in
its cache and $a_b(t) = 0$ leaves the cache unchanged.
When the cache is full and $a_b(t) = 1$ is selected,
the item with the lowest current utility $y^f_b(t)$ is
evicted to free the required space; if the freed space is
still insufficient, additional low-utility items are
removed until the incoming content can be
accommodated~\cite{niknia2025attention}.

\subsubsection{System Dynamics}

State transitions in the SMDP at BS $b$ are governed
jointly by the transition probability kernel
$P_{ss'}^b$ and the associated sojourn time $\tau_b$
between consecutive decision epochs.
The stochastic inter-arrival times introduced by the
Poisson request process make $\tau_b$ a random variable,
which is precisely the feature that distinguishes an SMDP
from a standard MDP~\cite{puterman2014markov}.
Because the system dynamics and reward structure are
generally unknown in practice, we employ reinforcement
learning to estimate them from experience and iteratively
refine the caching policy, as detailed in
Section~\ref{sec:algorithm}.

\subsubsection{Reward Function}

The instantaneous reward collected by BS $b$ at
decision epoch $t$ is

\begin{equation}
\begin{aligned}
r(t) = & \; w_1 \left[ \left( \mathbf{b}(t) \cdot \mathbf{d}(t) \right) \left( \mathbf{b}(t) \cdot \mathbf{y}(t) \right)^T  \; \right] \\
& \; - w_2 \left[ Mem(t) \right].
\end{aligned}
\label{eq:reward}
\end{equation}

where $w_1$ and $w_2$ are non-negative weights that
balance the two competing objectives, and $T$ is a
normalization constant. The first term rewards the BS for keeping content that
is simultaneously popular and high-utility in its cache:
$\mathbf{b}_b(t) \cdot \mathbf{d}_b(t)$ accumulates the
popularity of all currently cached items, while
$\mathbf{b}_b(t) \cdot \mathbf{y}_b(t)$ accumulates
their utilities, so the product is large only when the
cache holds fresh, important, and frequently requested
content.
The second term penalizes idle cache space: a large
$\mathrm{Mem}_b(t)$ signals that available storage is
being wasted, which the reward function discourages.

To illustrate, consider two scenarios at BS $b$ in which
all parameters are fixed except one.
In the first scenario, the cache is fully occupied with items that
are both fresh and frequently requested; the first term
is maximized and the penalty term vanishes, yielding a
high reward.
In the second scenario, the same items remain cached but
sufficient time has elapsed for their freshness and
therefore their utility to decline noticeably; the
first term decreases even though cache occupancy and
request patterns are unchanged, reflecting the preference
for up-to-date content.
As a third illustration, suppose some cache slots are
vacant even though the cached items are highly useful;
the penalty term then reduces the reward, incentivizing
the agent to fill available storage with beneficial
content.
Together, these terms guide the caching agent toward a
policy that maintains fresh, important, and popular
content in the cache while making efficient use of the
finite capacity $M$.

\subsubsection{Optimization Objective}

Because the caching problem is modeled as an
infinite-horizon SMDP (no terminal state exists while
the BS is operational), the goal of the caching agent
at BS $b$ is to find a stationary policy $\pi_b^*$ that
maximizes the long-run average reward:
\begin{equation}
  \pi_b^* \;=\;
  \arg\max_{\pi_b} \;
  \lim_{T_0 \to \infty}
  \frac{1}{T_0}
  \sum_{t=0}^{T_0}
  r_b(t),
  \label{eq:objective}
\end{equation}
which simultaneously maximizes the average utility of
cached content and minimizes unused cache capacity over
time.
Since the BSs are independent, solving~\eqref{eq:objective}
for each BS $b \in \mathcal{B}$ separately is equivalent
to the global multi-BS optimization, and no joint
optimization across BSs is required.

\section{Meta-Reinforcement Learning with Gradient-Clustered BS Sampling for Multi-BS Edge Caching}
\label{sec:algorithm}

In our multi-BS network, every BS $b \in \mathcal{B}$ runs an
independent caching agent that must learn a policy tailored
to its own local traffic characteristics, namely its request
rate $\lambda_b$ and content popularity distribution
$\mathbf{d}_b(t)$.
At each BS, we adopt PPO ~\cite{schulman2017proximal} as
the underlying reinforcement learning algorithm for making
caching decisions.
We then build on this per-BS PPO agent with a meta-learning
layer that enables knowledge to be shared efficiently across BSs, allowing each BS to adapt to its local environment more rapidly and effectively than learning an optimal policy independently from scratch.
. Furthermore, we introduce a gradient-based clustering mechanism
for selecting which BSs participate in each meta-update.
This section presents these three components in turn,
followed by a theoretical proof establishing why
gradient-clustered BS sampling accelerates convergence of
the meta-learning process relative to uniform random
sampling.

\subsection{Per-BS Caching Policy via Proximal Policy Optimization}
\label{subsec:ppo}

We rely on PPO as the local
reinforcement learning algorithm executed independently at
every BS. Before describing how knowledge is transferred
across BSs, we justify this choice and summarize the
mechanics of PPO as applied to the caching decision at a
single BS.

\subsubsection{Why PPO}

Two natural alternatives for the caching agent are DDPG and Double Deep
Q-Learning (DDQL). Neither is well suited to our setting.
DDPG is built for continuous control problems and requires
an action space that varies smoothly; the caching decision
at a BS, however, is inherently discrete (cache or do not
cache), so adopting DDPG would require artificial
discretization of an algorithm not designed for that
purpose. DDQL handles discrete actions natively but is known
to converge considerably more slowly than policy-gradient
methods such as PPO, which is problematic given that our BSs
must adapt quickly to shifting traffic conditions. PPO, by
contrast, operates directly on discrete action spaces and
uses a clipped surrogate objective that constrains the size
of each policy update, yielding both fast and stable
training even under fluctuations in request rates and content popularity distributions. For these reasons, PPO is the local learning
algorithm of choice at every BS in our framework.

\subsubsection{PPO Mechanics at a Single BS}

Each BS $b$ maintains its own policy network
$\pi_{\theta_b}$ and value network $V_{\theta_b}$, updated
using locally observed transitions
$(s_b(t), a_b(t), r_b(t), s_b(t{+}1), \tau_b)$, as defined
in Section~\ref{sec:formulation}.
The policy is updated by maximizing the clipped surrogate
objective. Let $\rho_b(\theta_b)$ denote the probability
ratio between the current and previous policies at BS $b$,
\begin{equation}
  \rho_b(\theta_b) \;=\;
  \frac{\pi_{\theta_b}(a_b(t) \mid s_b(t))}
       {\pi_{\theta_{b,\mathrm{old}}}(a_b(t) \mid s_b(t))}.
  \label{eq:prob_ratio}
\end{equation}
The clipped surrogate objective is then
\begin{equation}
\begin{split}
  L^{\mathrm{PPO}}_b(\theta_b) = \min \Big(
    & \rho_b(\theta_b)\, \hat{A}_b(t),\; \\
    & \mathrm{clip}\big(\rho_b(\theta_b), 1-\epsilon,\, 1+\epsilon\big)\, \hat{A}_b(t)
  \Big),
\end{split}
\label{eq:ppo_local}
\end{equation}
where $\pi_{\theta_{b,\mathrm{old}}}$ denotes the policy
prior to the current update, $\hat{A}_b(t)$ is the local
advantage estimate at BS~$b$, and~$\epsilon$ is the
clipping
parameter shared across all BSs. The clipping term restricts
how far the updated policy can drift from
$\pi_{\theta_{b,\mathrm{old}}}$ within a single update,
preventing destructive policy steps while still allowing
meaningful progress at each iteration.

The advantage estimate is computed as
\begin{equation}
  \hat{A}_b(t) = r_b(t) + \gamma^{\tau_b} V_{\theta_b}(s_b(t+1)) - V_{\theta_b}(s_b(t)),
  \label{eq:advantage_local}
\end{equation}
where $s_b(t+1)$ denotes the state of BS $b$ following the
transition. Because the underlying decision process is an
SMDP rather than a standard MDP (Section~\ref{sec:formulation}),
the discount factor is raised to the power of the sojourn
time $\tau_b$ rather than held fixed at one transition per
step; this ensures rewards realized after longer inter-request
intervals are discounted proportionally more than rewards
realized after short ones \cite{niknia2025transfer}.

The value network at BS $b$ is trained by minimizing the
squared discrepancy between its prediction and the
bootstrapped target return,
\begin{equation}
  L^{\mathrm{value}}_b(\theta_b) = \left(V_{\theta_b}(s_b(t)) - R_b(t)\right)^2,
  \label{eq:value_local}
\end{equation}
\begin{equation}
  R_b(t) = r_b(t) + \gamma^{\tau_b} V_{\theta_b}(s_b(t+1)).
  \label{eq:target_return_local}
\end{equation}
so that the predicted value of a state converges toward the
sum of the immediate reward and the discounted value of the
subsequent state.
Equations~\eqref{eq:ppo_local}--\eqref{eq:value_local}
together define the local update rule executed independently
by every BS at every decision epoch.

\subsection{Meta-Reinforcement Learning Across Base Stations}
\label{subsec:metarl}

Learning a caching policy from scratch at each new
BS requires extensive trial-and-error, during which the system
operates suboptimally and wastes computational resources. To
avoid this, we require a pretrained policy that is
sufficiently general to perform reasonably well across diverse
traffic conditions, namely different content popularity
distributions and varying request rates, and can serve as a
strong starting point for any new BS.
Rather than randomly initializing the neural network weights
as in learning from scratch, this pretrained policy provides
a warm initialization that captures the shared structure of
the caching problem. With only a few fine-tuning updates
adapted to the new BS's local traffic characteristics, the
policy can quickly achieve performance comparable to the
optimal policy trained specifically for that environment. To
obtain such a pretrained policy, we employ a meta-reinforcement
learning approach built on the MAML framework~\cite{finn2017model}. However, to make this
approach scalable to large networks with many heterogeneous
BSs, we modify the standard MAML BS sampling strategy: rather
than uniformly sampling BSs at each meta-iteration, we employ
a gradient-based clustering mechanism that clusters BSs by
similarity of their local gradients and samples BSs from
each cluster proportional to its size. This modification ensures that all traffic regimes
present in the network are represented at every meta-iteration,
reducing the variance of the meta-gradient estimator and
accelerating convergence of the shared initialization.

\subsubsection{Meta-Learning Objective}

The goal of meta-learning is to find a single shared policy
$\theta$ that can quickly adapt to any BS in the network.
The learning process has two loops: an inner loop and an
outer loop.

In the inner loop, for each BS $b$, the shared policy
$\theta$ is taken and updated for a few gradient steps using
the local traffic data of that BS. The result is a locally
adapted policy $\theta_b'$, which is simply $\theta$ after
being fine-tuned to the conditions of BS $b$:
\begin{equation}
  \theta_b' = \theta - \alpha \, \nabla_\theta \, \mathcal{L}_b(\theta),
  \label{eq:inner_loop}
\end{equation}
where $\alpha$ is the inner-loop learning rate and
$\mathcal{L}_b(\theta)$ is the local PPO loss at BS $b$.

After adaptation, $\theta_b'$ is tested on BS $b$ with no
further weight updates. The loss measured during this test
is the \emph{post-adaptation loss}, and it tells us how well
the adapted policy performs on BS $b$'s actual traffic
conditions.

The outer loop then updates the shared $\theta$ to minimize
the average post-adaptation loss across all BSs:
\begin{equation}
  F(\theta) = \min_{\theta} \;
  \frac{1}{|\mathcal{B}|}
  \sum_{b \in \mathcal{B}} \mathcal{L}_b(\theta_b').
  \label{eq:outer_loop}
\end{equation}

\subsubsection{Meta-Gradient Estimation}

The outer-loop update to $\theta$ requires the gradient of
the objective in~\eqref{eq:outer_loop} with respect to
$\theta$, referred to as the meta-gradient. In principle this
meta-gradient should be computed using all $N = |\mathcal{B}|$
BSs in the network. In practice, however, computing local
adaptation steps for every BS at every meta-iteration is
costly, particularly as $N$ grows large. We therefore
approximate the true meta-gradient using only a subset
$S \subset \mathcal{B}$ of $K \ll N$ BSs sampled at each
meta-iteration:
\begin{equation}
  \hat{G}(\theta) = \frac{1}{K} \sum_{b \in S} g_b(\theta),
  \qquad
  g_b(\theta) \stackrel{\mathrm{def}}{=} \nabla_\theta \mathcal{L}_b(\theta_b'),
  \label{eq:meta_grad_estimator}
\end{equation}
which is then used in the meta-parameter update
$\theta \leftarrow \theta - \eta \, \hat{G}(\theta)$, with
$\eta$ being the outer-loop learning rate. The central design
question is how the subset $S$ should be chosen at each
meta-iteration. The most direct approach is to draw $S$
uniformly at random from $\mathcal{B}$. However, a
gradient-aware sampling strategy yields a meta-gradient
estimator with substantially lower variance, leading to
faster convergence of the meta-learning process.

\subsection{Gradient-Based Clustering for BS Sampling}
\label{subsec:clustering_idea}
 
\subsubsection{Motivation}
 
BSs with similar traffic characteristics, comparable request
rates and similar popularity distributions, tend to produce
local gradients $g_b(\theta)$ that point in similar
directions, since their underlying loss landscapes
$\mathcal{L}_b(\theta)$ are themselves similar. Conversely,
BSs operating under very different traffic regimes (e.g., a
densely populated urban BS versus a lightly loaded suburban
BS) produce local gradients that diverge substantially.
A meta-gradient estimator built by sampling BSs uniformly at
random is, in effect, blind to this structure: it may by
chance select several BSs from the same traffic regime while
omitting others entirely, producing an estimate that is
unrepresentative of the true population gradient $G^*(\theta)$
and therefore exhibits high variance from one meta-iteration
to the next.
 
\subsubsection{Clustering Procedure}
 
To address this, we periodically group BSs according to the
similarity of their most recently computed local gradients,
and then sample the batch $S$ in a stratified manner
from these groups, ensuring that every traffic regime present
in the network is represented at each meta-iteration.
Concretely, upon every $\Delta_c$ meta-iterations, we
 
\begin{enumerate}
  \item collect the most recent local gradient $g_b(\theta)$
        computed at each BS $b \in \mathcal{B}$;
  \item normalize each gradient to unit norm,
        $\tilde{g}_b(\theta) = g_b(\theta)/\|g_b(\theta)\|$,
        and partition the $N$ BSs into $K$ clusters
        $C_1, \ldots, C_K$ by applying cosine-similarity-based
        grouping to the normalized gradient vectors
        $\{\tilde{g}_b(\theta)\}_{b \in \mathcal{B}}$, so that
        BSs whose gradients point in similar directions are
        grouped together. Normalizing before clustering is
        necessary because cosine similarity measures only the
        angle between two gradients and is invariant to their
        magnitude, whereas the variance
        decomposition 
        (Section~\ref{subsec:convergence_proof}) is stated in
        terms of Euclidean distance. For unit-norm vectors the
        two coincide exactly,
        $\|\tilde{g}_i(\theta) - \tilde{g}_j(\theta)\|^2
        = 2 - 2\cos\bigl(\tilde{g}_i(\theta), \tilde{g}_j(\theta)\bigr)$,
        so clustering on normalized gradients directly
        minimizes the within-cluster Euclidean variance
        $\sigma_k^2$ that governs the variance reduction proved
        in Theorem~3;
\item given a total sampling budget of $m$ BSs per
        meta-iteration, allocate $m_k = \lfloor m\, n_k/N \rceil$ draws to
        cluster $C_k$, where $n_k = |C_k|$, where $\lfloor \cdot \rceil$ denotes rounding to the nearest integer, ensuring $m_k \ge 1$ for every non-empty cluster and $\sum_k m_k = m$. Larger
        clusters receive proportionally more draws; at every
        meta-iteration occurring between two consecutive
        re-clustering events, draw $m_k$ BSs independently
        and uniformly from each cluster $C_k$,
        $k = 1, \ldots, K$, and form the BS batch as the
        cluster-size-weighted average of their local
        gradients,
        \begin{equation}
          \hat{G}_{\mathrm{clust}}(\theta)
          = \sum_{k=1}^{K} \frac{n_k}{N}\,
            \frac{1}{m_k}\sum_{j=1}^{m_k} g_{b_{k,j}}(\theta).
        \end{equation}
        When the clusters are of equal size and $m = K$,
        this allocation reduces to drawing exactly one BS
        per cluster.
\end{enumerate}

Re-clustering every $\Delta_c$ iterations, rather than at
every iteration, keeps the computational overhead of the
clustering step low while still allowing the cluster
assignment to track gradual changes in BS traffic patterns
over the course of training. Within a given clustering
interval, proportional sampling from every cluster guarantees
that the BS batch always spans the full diversity of
traffic regimes present in the network, weighted by how
prevalent each regime is, in contrast to uniform random
sampling, which provides no such guarantee.
\subsection{Convergence Analysis: Gradient-Clustered Sampling Versus Random Sampling}
\label{subsec:convergence_proof}
 
We now prove that gradient-clustered BS sampling converges
faster than uniform random sampling, following the
convergence framework of Fallah
\textit{et al.}~\cite{fallah2020convergence}, which
established the first theoretical guarantees for
gradient-based MAML on nonconvex objectives. The argument has
three steps. First, both sampling strategies produce unbiased
estimates of the meta-gradient (Lemma~1). Second, we derive
the exact variance of each estimator and prove that the
clustered estimator has strictly smaller variance, with the
gap given in closed form by the between-cluster variance of
the local gradients (Theorems~1--3). Substituting the reduced variance into that analysis shows
that the convergence guarantee of~\cite{fallah2020convergence}
carries over to clustered sampling with the variance term
strictly reduced (Proposition~1).
 
\subsubsection{Setup and Sampling Schemes}

Consider a network of $N = |\mathcal{B}|$ BSs. At a given
meta-iteration, each BS $b$ carries a local meta-gradient
$g_b(\theta) \in \mathbb{R}^{d}$, and the quantity both
strategies must estimate is the population meta-gradient
\begin{equation}
  G^*(\theta) = \frac{1}{N}\sum_{b \in \mathcal{B}} g_b(\theta).
  \label{eq:pop_grad}
\end{equation}
By~\eqref{eq:inner_loop}, $\theta_b'$ depends on $\theta$ both
directly and through the inner-loop gradient
$\nabla\mathcal{L}_b(\theta)$; applying the chain rule through
$\theta_b'$,
\begin{equation}
  g_b(\theta)
  \stackrel{\mathrm{def}}{=}
  \nabla_\theta\, \mathcal{L}_b(\theta_b')
  = \left(\frac{\partial \theta_b'}{\partial \theta}\right)^{\!\top}
    \nabla\mathcal{L}_b(\theta_b'),
  \label{eq:chain_rule_step}
\end{equation}
where
\begin{equation}
  \frac{\partial \theta_b'}{\partial \theta}
  = I - \alpha\nabla^2\mathcal{L}_b(\theta),
  \label{eq:jacobian_step}
\end{equation}
so that
\begin{equation}
  g_b(\theta)
  = \bigl(I - \alpha\nabla^2_\theta\mathcal{L}_b(\theta)\bigr)
    \nabla\mathcal{L}_b(\theta_b').
  \label{eq:gb_explicit}
\end{equation}

Since $F(\theta)$
in~\eqref{eq:outer_loop} averages $\mathcal{L}_b(\theta_b')$
over $b \in \mathcal{B}$, linearity of differentiation gives
\begin{equation}
\begin{split}
  \nabla F(\theta)
  &= \nabla\!\left(\frac{1}{N}\sum_{b \in \mathcal{B}} \mathcal{L}_b(\theta_b')\right)
  = \frac{1}{N}\sum_{b \in \mathcal{B}} \nabla\mathcal{L}_b(\theta_b') \\
  &= \frac{1}{N}\sum_{b \in \mathcal{B}} g_b(\theta)
  = G^*(\theta),
\end{split}
  \label{eq:gstar_equals_gradF}
\end{equation}

where $F(\theta)$ is the meta-loss of~\eqref{eq:outer_loop}.

Accordingly,
$G^*(\theta)$ and $\nabla F(\theta)$ are used interchangeably.
The clustering step partitions $\mathcal{B}$ into $K$ clusters
$C_1, \ldots, C_K$ by gradient similarity, with
$n_k = |C_k|$ and $\sum_{k=1}^{K} n_k = N$. Both strategies
are given the same sampling budget of $m$ BS evaluations per
meta-iteration, where $m \ge K$.
 
\emph{Random sampling.} The batch consists of $m$ BSs
$b_1, \ldots, b_m$ drawn independently and uniformly from
$\mathcal{B}$, and the meta-gradient is estimated by the
sample mean
\begin{equation}
  \hat{G}_{\mathrm{rand}}(\theta)
  = \frac{1}{m}\sum_{j=1}^{m} g_{b_j}(\theta).
  \label{eq:est_rand}
\end{equation}
 
\emph{Gradient-clustered sampling.} The budget is divided
among the clusters in proportion to their sizes,
\begin{equation}
  m_k = m\,\frac{n_k}{N},
  \label{eq:prop_alloc}
\end{equation}
so that larger clusters receive proportionally more draws;
we assume $m$ is chosen so that each $m_k$ is a positive
integer (in practice $m_k$ is rounded and at least one BS is
drawn per cluster). From each cluster $C_k$, $m_k$ BSs are
drawn independently and uniformly, and the estimator is the
cluster-size-weighted average
\begin{equation}
  \hat{G}_{\mathrm{clust}}(\theta)
  = \sum_{k=1}^{K} \frac{n_k}{N}\,
    \frac{1}{m_k}\sum_{j=1}^{m_k} g_{b_{k,j}}(\theta),
  \label{eq:est_clust}
\end{equation}
where $b_{k,j}$ denotes the $j$-th BS drawn from cluster
$C_k$. 
 
\subsubsection{Assumptions}
 
We adopt the assumption framework under which the convergence
of gradient-based MAML was first
established~\cite{fallah2020convergence}, stated here for the
finite BS population.
 
\noindent\textbf{A1 (\emph{Bounded initial gap}).}
\par\noindent
$F$ is bounded below, and
$\bar{\Delta} \stackrel{\mathrm{def}}{=}
F(\theta_0) - \inf_{\theta} F(\theta) < \infty$.
 
\noindent\textbf{A2 (\emph{Smooth local losses}).}
\par\noindent
Each local loss $\mathcal{L}_b$ is twice continuously
differentiable and $L$-smooth,
$\|\nabla\mathcal{L}_b(\theta) - \nabla\mathcal{L}_b(\theta')\|
\le L\|\theta - \theta'\|$ for all $\theta, \theta'$, which
holds whenever the policy and value networks use
Lipschitz-continuous activations on bounded
inputs~\cite{wu2022node}.
 
\noindent\textbf{A3 (\emph{Lipschitz Hessians}).}
\par\noindent
Each Hessian $\nabla^2\mathcal{L}_b$ is $\rho$-Lipschitz
continuous. Together with A2 and an inner-loop stepsize
$\alpha \in (0, 1/(6L)]$, this ensures that the meta-loss $F$
admits the smoothness surrogate and the accompanying
stochastic meta-stepsize rule
of~\cite[Lemmas~4.8 and~4.9]{fallah2020convergence}.
 
\noindent\textbf{A4 (\emph{Bounded BS-gradient dispersion}).}
\par\noindent
For all $\theta$,
$\frac{1}{N}\sum_{b \in \mathcal{B}}
\|g_b(\theta) - \nabla F(\theta)\|^2 \le \sigma^2$.
This is the finite-population analogue
of~\cite[Assumption~4.5]{fallah2020convergence}. Lemma~2 below
refines it as $\sigma^2 = \sigma_W^2 + \sigma_B^2$.
 
\noindent\textbf{A5 (\emph{Per-BS estimation noise}).}
\par\noindent
The stochastic meta-gradient $\hat{g}_b$ that a sampled BS
computes from its finite local rollouts satisfies
$\mathbb{E}[\hat{g}_b \mid b] = g_b(\theta)$ and
$\mathbb{E}\bigl[\|\hat{g}_b - g_b(\theta)\|^2 \mid b\bigr]
\le \tilde{\sigma}^2$, the analogue
of~\cite[Assumption~4.6]{fallah2020convergence}.
 
Following~\cite[Definition~4.1]{fallah2020convergence}, a
point $\theta_\epsilon$ is an \emph{$\epsilon$-approximate
first-order stationary point} ($\epsilon$-FOSP) if
$\mathbb{E}[\|\nabla F(\theta_\epsilon)\|] \le \epsilon$.
 
\subsubsection{Unbiasedness}
 
\noindent\textbf{Lemma 1 (\emph{Both estimators are unbiased}).}
\emph{Under either sampling scheme,
$\mathbb{E}[\hat{G}(\theta)] = G^*(\theta)$.}
 
\emph{Proof.} For random sampling, each draw $b_j$ is uniform
on $\mathcal{B}$, so
$\mathbb{E}[g_{b_j}(\theta)] = \frac{1}{N}\sum_b g_b(\theta) = G^*(\theta)$,
and averaging over $j$ preserves this. For clustered sampling,
each draw $b_{k,j}$ is uniform on $C_k$, so
$\mathbb{E}[g_{b_{k,j}}(\theta)] = \mu_k$, where
$\mu_k = \frac{1}{n_k}\sum_{i \in C_k} g_i(\theta)$ is the
cluster mean. Substituting into~\eqref{eq:est_clust},
\begin{equation}
  \mathbb{E}[\hat{G}_{\mathrm{clust}}(\theta)]
  = \sum_{k=1}^{K}\frac{n_k}{N}\,\mu_k
  = \frac{1}{N}\sum_{k=1}^{K}\sum_{i \in C_k} g_i(\theta)
  = G^*(\theta),
\end{equation}
where the last equality uses the fact that the clusters
together contain every BS exactly once.
$\rule{1.8mm}{1.8mm}$
 
\subsubsection{Variance Decomposition of the Gradient Population}
 
Since the gradients are vectors, we quantify the fluctuation
of an estimator by its total variance,
\begin{equation}
  \mathrm{Var}[\hat{G}(\theta)]
  \stackrel{\mathrm{def}}{=}
  \mathbb{E}\!\left[\bigl\|\hat{G}(\theta) - \mathbb{E}[\hat{G}(\theta)]\bigr\|^2\right],
  \label{eq:total_var}
\end{equation}
i.e. the expected squared distance of the estimate from its
mean, equal to the trace of its covariance matrix. Define the
within-cluster variance of cluster $C_k$ and the population
variance of the whole network as
\begin{equation}
  \sigma_k^2 = \frac{1}{n_k}\sum_{i \in C_k}\bigl\|g_i(\theta) - \mu_k\bigr\|^2,
  \label{eq:defs_var}
\end{equation}
\begin{equation}
  \sigma^2 = \frac{1}{N}\sum_{b \in \mathcal{B}}\bigl\|g_b(\theta) - G^*(\theta)\bigr\|^2 .
  \label{eq:defs_var2}
\end{equation}
 
\noindent\textbf{Lemma 2 (\emph{Exact ANOVA decomposition}).}
\emph{For any partition of $\mathcal{B}$ into clusters
$C_1, \ldots, C_K$,}
\begin{equation}
  \sigma^2 = \sigma_W^2 + \sigma_B^2,
  \label{eq:anova}
\end{equation}
\emph{where}
\begin{equation}
  \sigma_W^2 = \sum_{k=1}^{K}\frac{n_k}{N}\,\sigma_k^2,
  \qquad
  \sigma_B^2 = \sum_{k=1}^{K}\frac{n_k}{N}\,\bigl\|\mu_k - G^*(\theta)\bigr\|^2 .
  \label{eq:within_between}
\end{equation}

Here $\sigma_W^2$ is the size-weighted average of the
variances within the clusters and $\sigma_B^2$ is the
size-weighted variance of the cluster means around the
population mean.
 
\emph{Proof.} For $i \in C_k$, write
$g_i(\theta) - G^*(\theta) = \bigl(g_i(\theta) - \mu_k\bigr) + \bigl(\mu_k - G^*(\theta)\bigr)$
and expand the squared norm:
\begin{equation}
\begin{split}
  \bigl\|g_i(\theta) - G^*(\theta)\bigr\|^2
  ={}& \bigl\|g_i(\theta) - \mu_k\bigr\|^2
     + \bigl\|\mu_k - G^*(\theta)\bigr\|^2 \\
   & + 2\bigl\langle g_i(\theta) - \mu_k,\; \mu_k - G^*(\theta)\bigr\rangle .
\end{split}
\end{equation}
Summing over $i \in C_k$, the inner-product term vanishes.
To see why, note that $\mu_k - G^*(\theta)$ is the same vector
for every $i$ in the cluster, so it can be factored out of the
sum:
\begin{equation}
\begin{split}
  &\sum_{i \in C_k} 2\bigl\langle g_i(\theta) - \mu_k,\;
    \mu_k - G^*(\theta)\bigr\rangle \\
  &\quad = 2\Bigl\langle \sum_{i \in C_k}\bigl(g_i(\theta) - \mu_k\bigr),\;
    \mu_k - G^*(\theta)\Bigr\rangle.
\end{split}
\end{equation}
By the definition of $\mu_k$, the sum of deviations from the
cluster mean is exactly zero. Since
$\mu_k = \frac{1}{n_k}\sum_{i \in C_k} g_i(\theta)$,
we have $\sum_{i \in C_k} g_i(\theta) = n_k\mu_k$, and
therefore
$\sum_{i \in C_k}\bigl(g_i(\theta) - \mu_k\bigr)
= n_k\mu_k - n_k\mu_k = 0$,
since deviations above and below the mean always cancel.
The inner product with the zero vector is zero, so the entire
cross term vanishes. What remains after summing over
$i \in C_k$ is
\begin{equation}
\begin{split}
  \sum_{i \in C_k}\bigl\|g_i(\theta) - G^*(\theta)\bigr\|^2
  ={} & \sum_{i \in C_k}\bigl\|g_i(\theta) - \mu_k\bigr\|^2 \\
    & + \sum_{i \in C_k}\bigl\|\mu_k - G^*(\theta)\bigr\|^2 .
\end{split}
\end{equation}
For the first sum, recall from~\eqref{eq:defs_var} that
$\sigma_k^2 = \frac{1}{n_k}\sum_{i \in C_k}\|g_i(\theta) - \mu_k\|^2$,
so $\sum_{i \in C_k}\|g_i(\theta) - \mu_k\|^2 = n_k\sigma_k^2$.
For the second sum, $\|\mu_k - G^*(\theta)\|^2$ does not depend
on $i$, so summing the same quantity $n_k$ times gives
$n_k\|\mu_k - G^*(\theta)\|^2$. Therefore
\begin{equation}
  \sum_{i \in C_k}\bigl\|g_i(\theta) - G^*(\theta)\bigr\|^2
  = n_k\sigma_k^2 + n_k\bigl\|\mu_k - G^*(\theta)\bigr\|^2 .
\end{equation}
Summing over all $K$ clusters and dividing by $N$
gives~\eqref{eq:anova}--\eqref{eq:within_between}.
$\rule{1.8mm}{1.8mm}$
 
\subsubsection{\textbf{Theorem 1 (Variance of Random Sampling)}}
 
\emph{The random-sampling estimator~\eqref{eq:est_rand}
satisfies}
\begin{equation}
  \mathrm{Var}\!\left[\hat{G}_{\mathrm{rand}}(\theta)\right]
  = \frac{\sigma^2}{m}.
  \label{eq:var_rand}
\end{equation}
 
\emph{Proof.} The draws $b_1, \ldots, b_m$ are independent and
identically distributed, uniform on $\mathcal{B}$. A single
draw has mean $G^*(\theta)$ and total variance
$\mathbb{E}\bigl[\|g_{b_j}(\theta) - G^*(\theta)\|^2\bigr] = \sigma^2$
by~\eqref{eq:defs_var}. For independent vector-valued random
variables, the total variance of the sum equals the sum of the
total variances, and scaling by $1/m$ scales the variance by
$1/m^2$. Hence
\begin{equation}
  \mathrm{Var}\!\left[\hat{G}_{\mathrm{rand}}(\theta)\right]
  = \frac{1}{m^2}\sum_{j=1}^{m}\sigma^2
  = \frac{\sigma^2}{m}. \qquad \rule{1.8mm}{1.8mm}
\end{equation}
 
\subsubsection{\textbf{Theorem 2 (Variance of Gradient-Clustered Sampling)}}
 
\emph{The clustered-sampling estimator~\eqref{eq:est_clust}
with proportional allocation~\eqref{eq:prop_alloc} satisfies}
\begin{equation}
  \mathrm{Var}\!\left[\hat{G}_{\mathrm{clust}}(\theta)\right]
  = \frac{\sigma_W^2}{m}.
  \label{eq:var_clust}
\end{equation}
 
\emph{Proof.} All draws are independent, within and across
clusters. The draws from cluster $C_k$ are uniform on $C_k$,
so each has total variance $\sigma_k^2$, and their average
over $m_k$ draws has variance $\sigma_k^2 / m_k$. By
independence across clusters, the variance
of~\eqref{eq:est_clust} is the weighted sum
\begin{equation}
  \mathrm{Var}\!\left[\hat{G}_{\mathrm{clust}}(\theta)\right]
  = \sum_{k=1}^{K}\left(\frac{n_k}{N}\right)^{2}\frac{\sigma_k^2}{m_k}.
\end{equation}
Substituting the proportional allocation
$m_k = m\,n_k/N$ from~\eqref{eq:prop_alloc},
\begin{equation}
\begin{split}
  \mathrm{Var}\!\left[\hat{G}_{\mathrm{clust}}(\theta)\right]
  &= \sum_{k=1}^{K}\left(\frac{n_k}{N}\right)^{2}
    \frac{\sigma_k^2\,N}{m\,n_k} \\
  &= \frac{1}{m}\sum_{k=1}^{K}\frac{n_k}{N}\,\sigma_k^2
  = \frac{\sigma_W^2}{m},
\end{split}
\end{equation}
using the definition of $\sigma_W^2$
in~\eqref{eq:within_between}. $\rule{1.8mm}{1.8mm}$
 
\subsubsection{\textbf{Theorem 3 (Exact and Universal Variance Reduction)}}
 
\emph{For any cluster sizes $n_1, \ldots, n_K$ and any budget
$m$,}
\begin{equation}
  \mathrm{Var}\!\left[\hat{G}_{\mathrm{rand}}(\theta)\right]
  - \mathrm{Var}\!\left[\hat{G}_{\mathrm{clust}}(\theta)\right]
  = \frac{\sigma_B^2}{m} \;\ge\; 0,
  \label{eq:reduction}
\end{equation}
\emph{with strict inequality if and only if at least one
cluster mean $\mu_k$ differs from $G^*(\theta)$.}
 
\emph{Proof.} Subtracting~\eqref{eq:var_clust}
from~\eqref{eq:var_rand} and applying the ANOVA identity of
Lemma~2,
\begin{equation}
  \frac{\sigma^2}{m} - \frac{\sigma_W^2}{m}
  = \frac{\sigma^2 - \sigma_W^2}{m}
  = \frac{\sigma_B^2}{m}.
\end{equation}
Since $\sigma_B^2$ is a size-weighted sum of the non-negative
terms $\|\mu_k - G^*(\theta)\|^2$, it is non-negative, and it
equals zero if and only if every cluster mean coincides with
the population mean. $\rule{1.8mm}{1.8mm}$
 
Theorem~3 is exact: it involves no approximation, no
finite-population correction, and no assumption on the cluster
sizes. It also makes precise the role of the clustering
criterion. The reduction $\sigma_B^2/m$ is large exactly when
the cluster means are far apart, and grouping BSs by gradient
similarity is the mechanism that drives the cluster means
apart: BSs whose gradients point in similar directions are
absorbed into the same cluster, making each cluster internally
homogeneous (small $\sigma_k^2$, hence small $\sigma_W^2$) and
the clusters mutually distinct (large $\sigma_B^2$). By
Lemma~2 these two effects are two sides of the same identity,
since $\sigma_W^2 + \sigma_B^2 = \sigma^2$ is fixed by the
gradient population: every unit of variance the clustering
removes from within the clusters is transferred to the
between-cluster term, and by Theorem~3 that transferred
variance is precisely what the estimator no longer pays for.
 
\subsubsection{\textbf{Corollary 1 (Batch-to-Batch Stability)}}
 
The variance comparison of Theorem~3 admits a direct
interpretation in terms of how much the sampled BS batch
changes the meta-update from one meta-iteration to the next.
Let $\hat{G}^{(1)}(\theta)$ and $\hat{G}^{(2)}(\theta)$ denote
the meta-gradient estimates produced by two independent
samplings at the same iterate $\theta$. For any unbiased
estimator, the expected squared difference between two
independent realizations equals twice the variance,
\begin{equation}
  \mathbb{E}\!\left[\bigl\|\hat{G}^{(1)}(\theta) - \hat{G}^{(2)}(\theta)\bigr\|^2\right]
  = 2\,\mathrm{Var}\!\left[\hat{G}(\theta)\right],
  \label{eq:batch_diff}
\end{equation}
which follows by expanding the square and using the
independence of the two samplings. Substituting Theorems~1
and~2,
\begin{equation}
  \mathbb{E}\!\left[\bigl\|\hat{G}^{(1)}_{\mathrm{clust}} - \hat{G}^{(2)}_{\mathrm{clust}}\bigr\|^2\right]
  = \frac{2\sigma_W^2}{m},
  \label{eq:batch_stability_clust}
\end{equation}
\begin{equation}
  \mathbb{E}\!\left[\bigl\|\hat{G}^{(1)}_{\mathrm{rand}} - \hat{G}^{(2)}_{\mathrm{rand}}\bigr\|^2\right]
  = \frac{2\sigma^2}{m}.
  \label{eq:batch_stability_rand}
\end{equation}
Under clustered sampling, successive BS batches therefore
produce meta-updates that differ only by the within-cluster
term $2\sigma_W^2/m$: because every cluster is represented in
the same proportion at every meta-iteration, the composition
of the batch is essentially fixed, and consecutive updates
point in nearly the same direction. Under random sampling, the
cluster composition of the batch itself fluctuates, and
successive updates additionally differ by the between-cluster
term $2\sigma_B^2/m$, causing the update direction to swing
between meta-iterations. This stability is achieved without
sacrificing randomness: by Lemma~1, the clustered estimator
remains unbiased, and within each cluster the draw is
uniformly random, so every BS in the network continues to be
sampled over the course of training. Clustered sampling thus
retains the exploration benefits of stochastic BS selection
while removing the batch-composition fluctuation, and this reduction in the variability of the update direction is formally characterized in Proposition 1 below.
 
\subsubsection{\textbf{Proposition 1 ($\epsilon$-FOSP Complexity Under Clustered BS Sampling)}}

\emph{Let Assumptions A1--A5 hold, let
$\alpha \in (0, 1/(6L)]$, and let the meta-stepsize be chosen
by the rule of~\cite[Lemma~4.9]{fallah2020convergence}. Then
the guarantee of~\cite[Theorem~4.12]{fallah2020convergence}
for MAML applies unchanged to gradient-clustered proportional
BS sampling of budget $m$, with the BS-selection variance
term $\sigma^2/m$ replaced by $\sigma_W^2/m$: for any
$\epsilon \in (0,1)$, an $\epsilon$-FOSP $\theta_\epsilon$
satisfying}
\begin{equation}
  \mathbb{E}\bigl[\|\nabla F(\theta_\epsilon)\|\bigr]
  \;\le\;
  O\!\left(\sqrt{\frac{\sigma_W^2}{m} + \nu}\,\right) + \epsilon,
  \label{eq:fosp_clust}
\end{equation}
\emph{is found after at most $O(\bar{\Delta}L/\epsilon^2)$
meta-iterations, where $\nu$ collects the remaining per-BS
rollout-noise \footnote{Rollout noise
arises because each sampled BS estimates its local
meta-gradient from a finite number of environment
interactions rather than from the true expected gradient;
since this noise depends only on the rollout length at the
individually sampled BS and not on how the batch was
assembled, $\nu$ is independent of the BS-selection rule
and cancels out of any comparison between the two samplers.} and inner-loop-bias terms
of~\cite[Theorem~4.12]{fallah2020convergence}. Under uniform random
BS sampling, the same substitution recovers the original
guarantee, with $\sigma^2 = \sigma_W^2 + \sigma_B^2$ in place
of $\sigma_W^2$; since $\sigma_W^2 < \sigma^2$ whenever
$\sigma_B^2 > 0$ (Theorem~3), the accuracy attainable under
clustered sampling at any fixed budget $m$ is strictly better
than under uniform random sampling.}

\emph{Justification.} This is an application
of~\cite[Theorem~4.12]{fallah2020convergence} rather than an
independent result: the cited argument depends on the
BS-selection rule only through the second moment of the
batch-averaged per-BS meta-gradient about
$\nabla F(\theta_k)$, which for any unbiased sampler with
independent draws decomposes into a BS-selection variance
and the sampler-independent contribution $\nu$. The
gradient-clustered proportional sampler satisfies the three
properties this argument requires: unbiasedness (Lemma~1),
independence of the draws across and within clusters (by
construction), and a BS-selection variance of
$\sigma_W^2/m$ (Theorem~2); identically distributed draws are
not required, as the argument does not use this property. The
remaining ingredients of the cited proof, namely the
smoothness surrogate, the stochastic-stepsize moment bounds,
and the inner-loop bias control
of~\cite[Lemmas~4.8--4.10]{fallah2020convergence}, involve
only quantities local to the individually sampled BSs and are
therefore independent of the selection rule. Substituting
$\sigma_W^2/m$ for $\sigma^2/m$ yields the claim.
$\rule{1.8mm}{1.8mm}$

\subsubsection{Discussion}

Proposition~1 shows that gradient-clustered BS sampling
preserves the best-known iteration complexity
$O(\bar{\Delta}L/\epsilon^2)$ of MAML for nonconvex
objectives~\cite{fallah2020convergence} while strictly
reducing the floor to which the expected gradient norm
$\mathbb{E}[\|\nabla F(\theta_\epsilon)\|]$ can be driven
at any fixed sampling budget $m$.

The magnitude of this gain is governed by the
between-cluster variance $\sigma_B^2$. Three limiting cases
confirm the result is sharp: (i) If all BSs produce identical
gradients, then $\sigma_B^2 = 0$, both samplers coincide, and
the advantage vanishes, as it must when there is no diversity
to stratify over. (ii) If the clustering is uninformative, so that
every cluster mean happens to equal the population mean, then
again $\sigma_B^2 = 0$ and nothing is gained, which shows the
benefit is attributable to the quality of the clustering
rather than to stratification per se. (iii) In the regime
relevant to multi-BS edge caching, where BSs differ
substantially in request rate $\lambda_b$ and popularity
distribution $\mathbf{d}_b(t)$, the local losses
$\mathcal{L}_b$ and hence the local gradients differ markedly
across traffic regimes; gradient-similarity clustering
separates these regimes into distinct clusters with distinct
means, making $\sigma_B^2$ large. Two practical remarks
complete the picture. First, the guarantee at iteration $t$
is stated with respect to the cluster assignment in effect at
that iteration; between re-clustering events the assignment
may become stale as $\theta$ evolves, which can shrink the
realized $\sigma_B^2(\theta_t)$ but can never make clustered
sampling worse than random, since Theorems~1--3 hold for any
partition, and the re-clustering interval $\Delta_c$ thus
trades clustering overhead against staleness. Second, by
Corollary~1, the improvement manifests during training as
BS batches whose composition is essentially fixed across
meta-iterations, so consecutive meta-updates point in nearly
the same direction while every BS continues to be sampled
over time, yielding the more stable meta-training that
Proposition~1 quantifies.

The connection between the variance reduction and convergence
is made explicit through Proposition~1: reducing the
estimator variance from $\sigma^2/m$ to $\sigma_W^2/m$
directly lowers the irreducible floor
$O(\sqrt{\sigma^2/m+\nu})$ to $O(\sqrt{\sigma_W^2/m+\nu})$,
which is the lowest value of $\mathbb{E}[\|\nabla
F(\theta)\|]$ that any number of meta-iterations can achieve
at budget $m$. Both samplers reach this floor at the same
iteration rate $O(\bar{\Delta}L/\epsilon^2)$, but the floor
itself is strictly lower under clustered sampling whenever
$\sigma_B^2 > 0$. The variance analysis of Theorems~1--3
therefore does not merely characterize the estimator: it
directly determines how close to a stationary point of the
meta-loss the training process can converge.

To see explicitly how convergence is improved, note that the
bound in~\eqref{eq:fosp_clust} has two parts: the term
$\epsilon$, which can be made arbitrarily small by running
more meta-iterations (at a cost of $O(\bar{\Delta}L/\epsilon^2)$
iterations), and the irreducible floor
$O(\sqrt{\sigma_W^2/m + \nu})$, which cannot be reduced by
running longer regardless of how many iterations are used.
This floor is the fundamental limit imposed by the finite
sampling budget $m$ and the finite rollout length. Under
uniform random sampling the identical argument applies with
$\sigma^2$ in place of $\sigma_W^2$, giving a strictly higher
floor $O(\sqrt{\sigma^2/m + \nu})$. Since
$\sigma_W^2 < \sigma^2$ whenever $\sigma_B^2 > 0$
(Theorem~3), gradient-clustered sampling attains a lower
irreducible floor at the same budget $m$ and the same number
of iterations. In other words, both methods converge at the
same rate $O(\bar{\Delta}L/\epsilon^2)$, but they converge
to different limiting values of the expected gradient norm:
clustered sampling converges to a point closer to a stationary
point of the meta-loss than random sampling can reach at the
same cost. The gap between the two floors,
$O(\sqrt{\sigma^2/m+\nu}) - O(\sqrt{\sigma_W^2/m+\nu})$,
grows with $\sigma_B^2$, i.e., with the degree of
heterogeneity among the BSs, so the convergence advantage is
largest precisely in the heterogeneous multi-BS deployments
that motivate this work.

\section{Experimental Setup and Results}
\label{results}

This section outlines the experimental setup used to evaluate
the proposed gradient-based clustered meta-RL caching framework,
followed by the baseline methods and the evaluation criteria.
The multi-BS environment and all learning agents were
implemented in Python~3 using the TensorFlow
framework~\cite{abadi2016tensorflow}.

\subsection{Configuration and Parameters}

We simulate a network of $N = 62$ BSs, of which $60$ are used
for meta-training and $2$ are held out as unseen test
environments. Heterogeneity is induced along the two axes discussed
in Section~\ref{sec:sysmodel}: each BS $b$ is assigned its own
Zipf popularity skewness $\zeta_b \in [0.1, 3.0]$ and its own
mean request inter-arrival time
$1/\lambda_b \in [0.05, 2.0]$~s, with arrivals following an
independent Poisson process of rate
$\lambda_b$~\cite{gomaa2013estimating}. 

Each BS serves $F = 50$ content types, and its cache have a capacity of $M = 10{,}000$ storage
units. Content lifetimes, sizes, and importance values are
drawn from $\{1, 10, 15, 20, 25, 30\}$,
$\{100, 200, 700, 1000\}$, and $\{0.3, 0.5, 0.7, 0.9\}$,
respectivelys. The utility function $\mathcal{E}(\cdot)$
of~\eqref{ufunc} grows linearly with importance and decays
exponentially with age, following~\cite{niknia2025attention}.

The local PPO updates
of~\eqref{eq:ppo_local}--\eqref{eq:value_local} use discount
factor $\gamma = 0.99$, and
minibatch size $64$.

For the meta-learning loop of Section~\ref{subsec:metarl},
each sampled BS performs one inner-loop update
of~\eqref{eq:inner_loop} with $\alpha = 10^{-3}$ on a support
rollout of $200$ transitions, and its local meta-gradient is
evaluated on a disjoint query rollout of $100$ transitions.
The meta-parameters are updated with Adam at outer-loop rate
$\eta = 10^{-4}$. Following Section~\ref{subsec:clustering_idea}, the BSs are
partitioned into $K = 6$ clusters by applying $k$-means to the
normalized local gradients $\tilde{g}_b(\theta)$, so that BSs
are grouped by the cosine similarity of their gradients, and
the partition is refreshed every $\Delta_c = 10$
meta-iterations from the most recently observed gradients;
before any gradient is available, the BSs are partitioned
uniformly at random. The sampling budget is $m = 10$ BS
evaluations per meta-iteration. 

\renewcommand{\arraystretch}{1.3}
\begin{table}[htbp]
  \centering
  \caption{Parameter settings}
  \label{table:params}
  \begin{tabular}{lp{0.45\linewidth}}
    \hline
    \textbf{Notation} & \textbf{Value} \\
    \hline
    $N$ & 62 (2 held out) \\
    $F$ & 50 \\
    $M$ & 10{,}000 \\
    $\zeta_b$ & $[0.1, 3.0]$ \\
    $1/\lambda_b$ & $[0.05, 2.0]$ \\
    $l_f$ & $\{1, 10, 15, 20, 25, 30\}$ \\
    $z_f$ & $\{100, 200, 700, 1000\}$ \\
    $i_f$ & $\{0.3, 0.5, 0.7, 0.9\}$ \\
    $\gamma$ & 0.99 \\
    Batch size & 64 \\
    $\alpha$ & $10^{-3}$ \\
    Inner-loop steps & 1 \\
    $\eta$ & $10^{-4}$ \\
    Support / query rollout & 200 / 100 transitions \\
    $K$ & 6 \\
    $\Delta_c$ & 10 \\
    \hline
  \end{tabular}
\end{table}

\subsection{Baselines}

To separate the contribution of gradient-clustered BS
sampling from that of meta-learning itself, we compare against
the following baselines. All methods share the same environment,
the per-BS SMDP formulation of Section~\ref{sec:formulation},
the network architecture, and the hyperparameters of
Table~\ref{table:params}; they differ only in how, and
whether, knowledge is shared across BSs.

\begin{itemize}
  \item \textbf{MAML with Random Sampling (MAML-RS):} a meta-RL caching baseline adapted from the cooperative meta-RL scheme of~\cite{wei2024cooperative}. Whereas the original method operates in a multi-agent setting with cooperation among nodes, we implement it here without the multi-agent component, consistent with our system model in which BSs are independent, so that the meta-training BS batch is selected by uniform random sampling. Comparing against it isolates the benefit of replacing random BS selection with the proposed gradient-clustered proportional sampler.

  \item \textbf{Learning from Scratch (LfS):} the policy at
  the target BS is trained with randomly initialized actor and
  critic networks and no prior knowledge \cite{schulman2017proximal}.

  \item \textbf{Transfer Learning (TL):} a baseline adapted from
  our previous work~\cite{niknia2025transfer}. Rather than
  transferring demonstration data from the source environment
  as in~\cite{niknia2025transfer}, here we transfer only the actor
  and critic networks trained to convergence at a single source
  BS to initialize the networks at the target BS, which are
  then fine-tuned on the target's local traffic. This isolates
  the impact of the neural network initialization alone on
  performance in the new environment.

  \item \textbf{Meta-learning with Gradient-Clustered Sampling (MGCS):} our proposed approach, in which the shared meta-policy is trained with the MAML outer loop while the BS batch at each meta-iteration is selected by the gradient-clustered proportional sampler of Section~\ref{subsec:clustering_idea}, rather than uniformly at
random.
\end{itemize}

\subsection{Evaluation Criteria}

The performance of the proposed approach is assessed using the
following criteria.

\begin{itemize}
  \item \textbf{Meta-training convergence:} the outer-loop
  meta-loss of~\eqref{eq:outer_loop}, averaged over the sampled
  BS batch, as a function of meta-iterations. A method that
  attains a lower meta-loss at the same iteration count, or
  reaches a given level in fewer iterations, converges faster.

  \item \textbf{Empirical meta-gradient variance:} the sample
  variance of the meta-gradient estimator across
  meta-iterations at a fixed iterate, reported for the
  clustered and the uniform random sampler at equal budget $m$.
  This validates the predicted variance gap empirically
  $\sigma_B^2/m$ of Theorem~3 and the batch-to-batch stability
  of Corollary~1.

  \item \textbf{Adaptation on held-out BSs:} two BSs are held out from meta-training and reserved for evaluating adaptation to unseen environments.
  The post-adaptation loss and average reward attained at the
  held-out BSs, whose traffic configurations are never used
  during meta-training, reported as a function of the number of
  local adaptation steps. This measures the quality of the
  learned initialization, which is the operational goal of the
  meta-learning framework of Section~\ref{subsec:metarl}.

  \item \textbf{Average reward:} for each training trial, the
  average of all instantaneous rewards~\eqref{eq:reward}
  accumulated up to that point, consistent with the long-run
  objective~\eqref{eq:objective}. The goal of the reinforcement
  learning agent is to maximize the average of the
  instantaneous rewards over the long run, so an approach that
  converges to a higher average reward yields a better caching
  policy. Since a higher reward corresponds to a higher cache
  hit rate, an algorithm achieving a higher long-term average
  reward demonstrates superior caching performance.

  \item \textbf{Convergence of the average reward:} the
  stability of the average-reward curve over training. In
  reinforcement learning, convergence is essential: a curve
  that keeps oscillating or trends downward indicates that the
  agent has failed to settle on a stable policy, whereas a
  curve that flattens and remains steady indicates that a
  consistent policy has been learned. An approach whose average
  reward converges to a stable value is therefore preferable to
  one whose reward continues to fluctuate or decline.

\end{itemize}

\subsection{Results}

\subsubsection{Meta-Training Convergence}

\begin{figure*}[!t]
  \centering
  \includegraphics[width=0.7\textwidth]{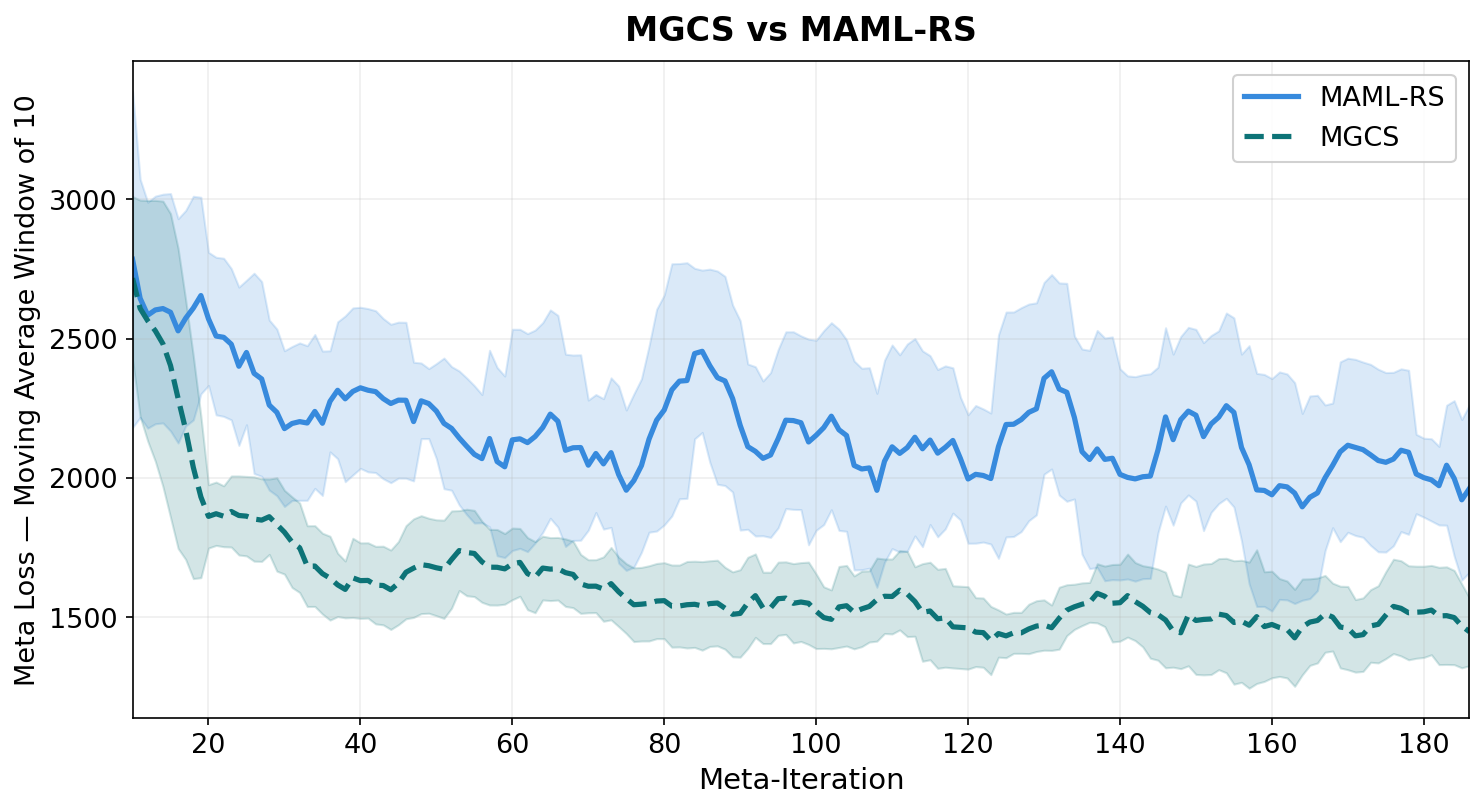}
  \caption{Meta-training loss of gradient-clustered sampling
  versus uniform random sampling. Curves show the moving
  average of the outer-loop meta-loss over a window of ten
  meta-iterations; shaded regions denote one standard
  deviation within the window. Gradient-clustered sampling
  converges faster, reaches a lower meta-loss, and exhibits
  markedly smaller iteration-to-iteration fluctuation.}
  \label{fig:meta_loss}
\end{figure*}

Fig.~\ref{fig:meta_loss} compares the meta-training loss of
the proposed gradient-clustered sampler against uniform random
sampling over the course of meta-training. Both methods use an
identical environment, network architecture, sampling budget
$m$, and set of hyperparameters, and differ only in how the
BS batch is selected at each meta-iteration; each curve
shows the moving average of the outer-loop
meta-loss~\eqref{eq:outer_loop} over a window of ten
iterations, and the shaded region denotes the standard
deviation of the meta-loss within that window.

Two effects predicted by the convergence analysis of
Section~\ref{subsec:convergence_proof} are visible. First,
gradient-clustered sampling converges faster and to a lower
meta-loss: it drops below a meta-loss of $1800$ within the
first $20$ meta-iterations, whereas random sampling requires
substantially longer and stabilizes at a markedly higher
level, with final values of $1447 \pm 122$ and $1961 \pm 300$,
respectively, a reduction of roughly $26\%$ in the converged
meta-loss. Second, the meta-loss of the clustered sampler
fluctuates far less from iteration to iteration: its shaded
band is consistently narrower than that of random sampling
across the entire training run. This reduced fluctuation is
the empirical signature of the lower meta-gradient variance
established in Theorem~3, whereby drawing a size-proportional
number of BSs from every cluster removes the between-cluster
component $\sigma_B^2$ from the variance of the meta-gradient
estimator; the update direction therefore varies less between
consecutive iterations, in agreement with the batch-to-batch
stability of Corollary~1.

The two effects reinforce each other. Under random sampling,
the BS batch composition swings from one meta-iteration to
the next as different BSs enter and leave the batch, so the
meta-gradient estimate is noisier and progress is slower and
less monotone, as reflected in the large oscillations of the
random-sampling curve. Under gradient-clustered sampling, the
batch always spans the full range of traffic regimes in
proportion to their prevalence, which yields a more accurate
and more stable descent direction and, in turn, a faster and
smoother convergence observed in Fig.~\ref{fig:meta_loss}.

\subsubsection{Adaptation to an Unseen Base Station}

We evaluate adaptation on two held-out BSs that were not seen
during meta-training, chosen to span the difficulty range of
the target task: one whose traffic makes the caching problem
relatively easy, and one whose traffic makes it considerably
harder. In both cases, the meta-learning methods deploy their
learned initialization and fine-tune on the held-out BS's
local traffic, TL is initialized from a source BS and
fine-tuned, and LfS is trained from a random initialization.

\paragraph{Easy BS}

\begin{figure}[!t]
  \centering
\includegraphics[width=\linewidth]{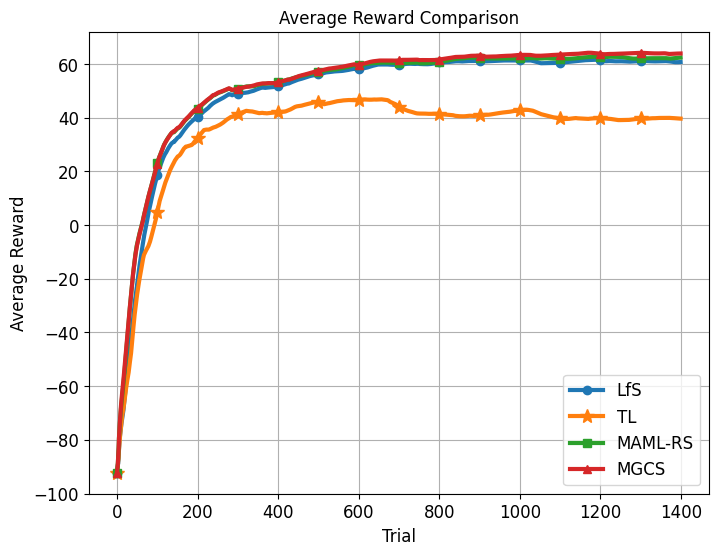}
  \caption{Average reward during adaptation to a held-out BS
  with $\zeta_b = 1$ and $\lambda_b = 5$. MGCS and MAML-RS
  overlap and reach the highest reward, confirming that
  gradient-clustered sampling preserves adaptability; both
  meta-learning methods adapt faster than LfS; and TL
  converges to a lower reward due to negative transfer from a
  dissimilar source ($\zeta = 0.22$, $\lambda = 3.22$).}
  \label{fig:reward_heldout}
\end{figure}

Fig.~\ref{fig:reward_heldout} reports the average reward of the
four methods when adapting to a held-out BS whose traffic
configuration, a Zipf skewness of $\zeta_b = 1$ and a request
rate of $\lambda_b = 5$, was not seen during meta-training.
This configuration
makes the caching problem relatively easy: the high Zipf
skewness concentrates demand on a small set of files, so the
most popular files are requested far more often than the rest,
and the shorter time between consecutive requests  means cached files remain valid
longer before they expire. The agent therefore faces little
ambiguity about which files to cache and when, and a good
caching policy is straightforward to learn.

Three observations follow. First, MGCS and MAML-RS are
essentially indistinguishable throughout adaptation, both
converging to the highest reward of roughly $64$ at a similar
rate. This is the intended outcome: gradient-clustered
sampling changes only \emph{which} BSs are drawn during
meta-training and leaves the meta-objective unchanged, so it is
expected to reduce meta-gradient variance without altering the
quality of the learned initialization.

Second, LfS performs close to both meta-learning methods,
adapting at a similar rate and reaching a comparable final
reward. Because adaptation here involves a single stationary
BS, the task is relatively easy, so a from-scratch agent can
learn a reasonable policy from a random initialization without
the benefit of meta-training, which is why its curve stays
close to those of MGCS and MAML-RS.

Third, TL settles at a substantially lower reward than the
other methods, plateauing near $40$ while the others exceed
$60$, and its reward curve declines over the course of
adaptation rather than rising toward a stable value. A falling
reward curve is a clear signature of negative transfer:
instead of building on the transferred initialization, the
target agent is driven away from a good policy, so that
fine-tuning degrades performance rather than improving it.
This illustrates that the effectiveness of transfer learning
depends strongly on the similarity between the source and
target environments. Here the source BS has $\zeta = 0.22$ and
$\lambda = 3.22$, whereas the target has $\zeta = 1$ and
$\lambda = 5$; the large gap between the two traffic regimes
causes the transferred initialization to steer the target
agent toward a policy suited to the source rather than the
target, so the transferred knowledge actively harms
adaptation. Negative transfer is the principal failure mode of
transfer learning and is difficult to predict in advance,
since it requires knowing the target environment before
committing to a source. Meta-learning avoids this pitfall by
construction: rather than transferring from one fixed source,
it learns an initialization optimized for adaptation across
the whole population of traffic regimes, so its performance
does not depend on any single source-target pairing.

Table~\ref{table:hitrate} reports the number of cache hits
for 1000 requests for held-out BSs after adaptation. Consistent with the
reward curves for easy task (Fig. ~\ref{fig:reward_heldout}), MGCS, MAML-RS, and LfS all perform comparably
(607, 593, and 596 hits for the easy task). Because this task is easy, a from-scratch agent
reaches essentially the same caching quality as the
meta-learning methods, and MGCS and MAML-RS remain
indistinguishable. TL again trails the other methods (550
hits), consistent with the negative transfer observed in its
reward curve.

\begin{figure}[!t]
  \centering
  \includegraphics[width=\linewidth]{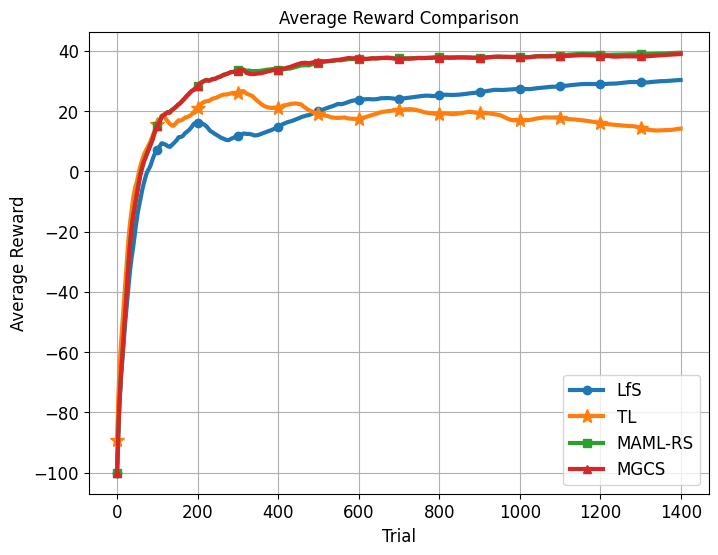}
  \caption{Average reward during adaptation to a difficult
  held-out BS with $\zeta_b = 0.22$ and $\lambda_b = 4.54$.
  MGCS and MAML-RS overlap and adapt fastest; LfS lags
  markedly because the low-skewness traffic makes the task
  hard to learn from scratch; and TL declines under negative
  transfer.}
  \label{fig:reward_hard}
\end{figure}

\paragraph{Difficult BS}
Fig.~\ref{fig:reward_hard} reports adaptation to a held-out BS
with $\zeta_b = 0.22$ and $\lambda_b = 1.13$, whose traffic
makes the caching problem substantially harder. The small Zipf
skewness means the popularities of the files are much closer to
one another, so there is no small set of clearly popular files
to prioritize and the agent has only a weak signal to decide
what to cache. In addition, the longer time between consecutive
requests means cached files are more likely to expire before
they are requested again, so the agent must also decide
carefully when to cache each file. Together, these two factors
make the caching decision considerably more difficult than on
the easy target. Here the value of a meta-trained
initialization becomes clear. MGCS and MAML-RS again overlap
and adapt fastest, reaching the highest reward of roughly $39$,
confirming once more that gradient-clustered sampling preserves
adaptability regardless of task difficulty. LfS, however, now
lags markedly behind: from a random initialization it improves
only slowly and plateaus around $30$, well below the
meta-learning methods, because on this harder task a
from-scratch agent cannot quickly discover a good policy from
local interaction alone. The gap between the meta-learning
methods and LfS, which was negligible on the easy target, is
now substantial and persists throughout adaptation, showing
that the benefit of a meta-trained initialization grows with
task difficulty. TL again fails to converge, its reward
declining over the course of adaptation and ending lowest of
all methods near $15$, a further instance of negative transfer.

\begin{table*}[!t]
  \centering
  \caption{Cache hits after adaptation at the easy and difficult held-out BSs}
  \label{table:hitrate}
  \renewcommand{\arraystretch}{1.3}
  \begin{tabular}{lcc}
    \hline
    \textbf{Method} & \textbf{Easy BS ($\zeta_b=1$, $\lambda_b=5$)} & \textbf{Difficult BS ($\zeta_b=0.22$, $\lambda_b=1.13$)} \\
    \hline
    TL      & 550 & 294 \\
    LfS     & 596 & 333 \\
    MAML-RS & 593 & 367 \\
    MGCS    & 607 & 366 \\
    \hline
  \end{tabular}
\end{table*}

In Table~\ref{table:hitrate},
MGCS and MAML-RS achieve the highest and essentially identical
performance (366 and 367 hits) for the difficult task, confirming that
gradient-clustered sampling matches the caching quality of
uniform random sampling. Both
meta-learning methods outperform LfS (333 hits), and all three
substantially exceed TL (294 hits), which is degraded by the
negative transfer discussed above.

The near-identical performance of MGCS and MAML-RS in
Figs.~\ref{fig:reward_heldout} and ~\ref{fig:reward_hard} confirms
that gradient-clustered sampling remains unbiased, as does
uniform random sampling, and therefore produces a meta-trained
initialization of comparable quality. At the same time, MGCS
achieves this performance with substantially fewer training
steps, demonstrating the training-efficiency benefit of
gradient-clustered sampling.

\section{Conclusion}
\label{Conclusion}

This paper addressed the high-variance meta-gradient estimation
that slows meta-training when a shared caching policy is learned
across many independent, heterogeneous BSs. We
proposed a meta-reinforcement learning framework in which each
BS runs a local PPO agent while a shared meta-policy is trained
with a MAML-style outer loop whose BS batch is selected by
gradient-based clustering, drawing from each cluster in
proportion to its size. Through an ANOVA-style decomposition of
the gradient variance, we proved that this sampler yields a
strictly lower-variance meta-gradient estimator than uniform
random sampling under BS heterogeneity, with the gap given in
closed form by the between-cluster variance, and showed that the
known MAML convergence guarantee still holds with the total
variance replaced by its smaller within-cluster part. Extensive
simulations confirmed the predicted variance reduction, showed
faster and more stable meta-training than random sampling, and
demonstrated that the resulting meta-policy adapts to unseen BSs
as effectively as the strongest baselines while avoiding the
negative transfer that degrades transfer learning. Several directions are left for future investigation. First,
this work fixes the number of clusters and the
re-clustering interval as treated hyperparameters; a
principled method for selecting them adaptively based on the
observed gradient dispersion could be studied in future work. Second, the
variance-reduction principle underlying gradient-clustered
sampling has only been applied here to edge caching; whether
it yields similar benefits in other heterogeneous wireless
network optimization tasks, such as resource allocation or
handover management, could be further explored.

\bibliographystyle{IEEEtran}
\bibliography{Bibliography}

\end{document}